\documentclass[journal]{IEEEtran}
\usepackage{amssymb}
\usepackage{amsmath}
\newcommand{\norm}[1]{\left\lVert#1\right\rVert}
\usepackage{subcaption}
\usepackage{rotating}
\usepackage{algorithmic}
\usepackage{algorithm}

\usepackage{multirow}
\usepackage{color}
\usepackage{xcolor}
\usepackage{boldline}
\usepackage{cellspace}
\usepackage{url}
\usepackage[english]{babel}
\usepackage{balance}
\usepackage{graphicx}
\begin{document}


\title{ADIC: Anomaly Detection Integrated Circuit in 65nm CMOS utilizing Approximate Computing}

\author{\IEEEauthorblockN{
Bapi Kar*,~\IEEEmembership{Member,~IEEE},
Pradeep Kumar Gopalakrishnan*,~\IEEEmembership{Senior Member,~IEEE},\\
Sumon Kumar Bose*,~\IEEEmembership{Student Member,~IEEE},
Mohendra Roy,~\IEEEmembership{Member,~IEEE}, and \\
Arindam Basu,~\IEEEmembership{Senior Member,~IEEE}
}\\
\thanks{This work was supported by Delta Electronics Inc. and the National Research Foundation Singapore under the Corp Lab @University scheme.}
\thanks{B. Kar is associated with Mentor Graphics, Bengaluru –560103, India (e-mail: bapik@ieee.org).}
\thanks{S. K. Bose, P. K. Gopalakrishnan and A. Basu are with the Delta-NTU Corporate Laboratory for Cyber-Physical Systems, School of EEE, NTU, Singapore 639798.  (e-mail: bose0003@e.ntu.edu.sg; PRAD0015@e.ntu.edu.sg; arindam.basu@ntu.edu.sg).}
\thanks{M. Roy is with Pandit Deendayal Petroleum University, Gandhinagar, India 382007 (e-mail:
mohendra.roy@sot.pdpu.ac.in).}
\thanks{*equal contribution.}
}

\maketitle

\begin{abstract}
In this paper, we present a low-power anomaly detection integrated circuit (ADIC) based on a one-class classifier (OCC) neural network. The ADIC achieves low-power operation through a combination of (a) careful choice of algorithm for online learning and (b) approximate computing techniques to lower average energy. In particular, online pseudoinverse update method (OPIUM) is used to train a randomized neural network for quick and resource efficient learning. An additional $42\%$ energy saving can be achieved when a lighter version of OPIUM method is used for training with the same number of data samples lead to no significant compromise on the quality of inference. 

Instead of a single classifier with large number of neurons, an ensemble of $K$ base learner approach is chosen to reduce learning memory by a factor of $K$. This also enables approximate computing by dynamically varying the neural network size based on anomaly detection. Fabricated in 65nm CMOS, the ADIC has $K=7$ Base Learners (BL) with $32$ neurons in each BL and dissipates $11.87 pJ/OP$ and $3.35 pJ/OP$ during learning and inference respectively at $V_{dd}=0.75V$ when all $7$ BLs are enabled. Further, evaluated on the NASA bearing dataset, approximately $80\%$ of the chip can be shut down for $99\%$ of the lifetime leading to an energy efficiency of $0.48 pJ/OP$, an $18.5$ times reduction over full-precision computing running at $V_{dd}=1.2V$ throughout the lifetime.
\end{abstract}

\begin{IEEEkeywords}
Anomaly Detection, Internet of Things, Edge computing, Predictive Maintenance, One Class Classification, Energy Savings, Approximate Computing.
\end{IEEEkeywords}

\section{Introduction}
\begin{figure}[!ht]
\centering
\includegraphics[scale=0.16]{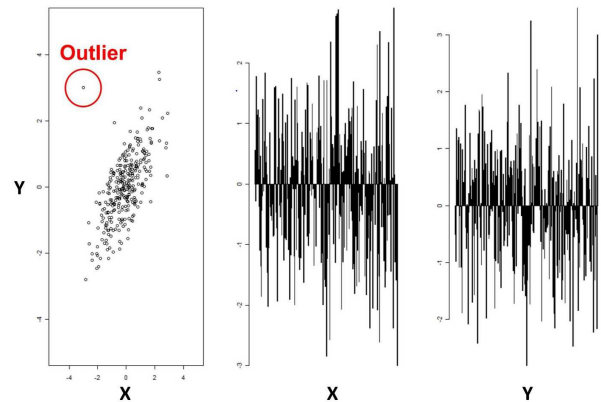}
\caption{Anomaly detection for two variables \cite{adet_1}}
\label{fig_ad_e1}
\end{figure}

Anomaly detection is an important topic in the field of machine learning and deep learning, which focuses on the efficient detection of anomalous activity in a process. A typical data visualization of an anomaly (outlier) is presented in Fig. \ref{fig_ad_e1}, for a two-dimensional ($2D$) data ($X$, $Y$) \cite{adet_1}. This $2D$ data is easy to visualize and requires relatively simpler statistical analysis to identify the highlighted outlier. For high dimensional data beyond $2D$, it requires more sophisticated computations like \textit{Mahalanobis distance}.  Typically, there is no prior knowledge to characterize an anomaly, nor ample data is available in any of the practical fields that symbolize these outliers \cite{adet_8}. A few examples of anomaly detection are the detection of rare (unusual) events in real life scenarios such as medical complications depicted in ECG/EEG data abnormalities \cite{adet_2, adet_3, adet_4}, malfunctioning of rotary machines due to various physical factors \cite{adet_5, adet_7}, security threat in cyber-physical systems \cite{adet_6, adet_10, adet_11, adet_12} or even fraud detection in financial systems \cite{adet_9}. 

This paper focuses on finding anomalies in rotating machines, also known as machine health monitoring in Industry 4.0. In many industrial setups, rotating machines are of the utmost of importance in its functioning. Their failure can lead to a catastrophic impact such as loss of productivity and accidents leading to safety concerns of the working personnel. Therefore, maintenance of these machines becomes very critical, thus requiring constant human supervision or by automatic instrumentation. Periodic maintenance based on statistical models and reliability data \cite{BRYAN} results in fewer breakdowns. However, this is not a cost-effective approach due to unacceptable wastage of the residual useful life of the healthy machines \cite{Kadry:2012:DPE:2432269}. Contrarily, reactive maintenance comes into effect only when a machine fails leading to an unpredictable loss in productivity. Therefore, the industry is poised to move towards Predictive Maintenance (PdM) and is shown to be more effective in reducing the downtime of a machine and the overall cost of maintenance. 

Machine learning (ML) based techniques in PdM such as aero-engine control system sensor fault detection \cite{Vishwanath}, fault diagnostics of rotary machines \cite{SANZ2007981} and machine health monitoring tasks \cite{khan_madden_2014} are a few examples of PdM which rely on detecting a trend of fault in a machine as an anomaly. Lack of adequate failure data prohibits a multi-class classification based fault detection model \cite{khan_madden_2014}. Furthermore, the failure signatures vary drastically from one machine to another and hence is not practical to build a generic detection system using a generic set of data pertaining to a set of machines of the same category. Instead, each model needs to learn the parameters from the sensor data attached to the machine corresponding to it. In summary, this is a classic case of \textit{One-Class Classification} (OCC), where a classifier is trained to learn the similar data distribution. In this work, the proposed anomaly detector model learns from the healthy data and identifies the occurrence of signatures corresponding to the mechanical failures as the deviations from a healthy state of the machines \cite{7727444}.

In a large enterprise, a large number of sensors are interconnected through a network to a central server, typically over the Internet, and are commonly referred to as \textit{Internet-Of-Things} (IoT). It involves a large amount of data transmission for subsequent processing over the network leading to significant power consumption and bandwidth requirement. This turns out to be a bottleneck in case of critical low latency remedial actions. Alternatively, the data processing engine is being pushed to the edge of the network, i.e., in the vicinity of the sensor modules, enabling the system to transmit the decision to the servers only, thus saving on energy consumption and reducing bandwidth requirement. In many IoT applications, the computing engines placed near the sensors need to be battery operated. This demands an edge computing device to be more energy efficient during its operation, thus elongating its operation lifetime \cite{8123913}. Our study showed that no low power integrated circuit (IC) implementation of an edge computing device exists for anomaly detection.

This paper presents a novel energy-efficient machine learning co-processor referred to as \textit{Anomaly Detection Integrated Circuit} (ADIC). This design implements a popular machine learning framework called \textit{Extreme Machine Learning} (ELM) \cite{HUANG2006489} used for shallow neural networks. The  key  differentiators of the proposed ASIC implementation from the microprocessor-based anomaly detector~\cite{adepos_j} are the followings:
\begin{enumerate}
\item Use of pseudo-random binary sequence (PRBS) based weight generation eliminating memory required for weights in the first stage of ELM.
\item Online learning using OPIUM \cite{Tapson2013} and OPIUM-Lite~\cite{VanSchaik2015a} learning rule (without matrix inverse) for individual devices to learn their own baseline condition.
\end{enumerate}
In addition, ADIC also supports ensemble classifier architecture to implement approximate computing using ADEPOS algorithm \cite{Bose:ASPDAC2019}.


In this paper, we start with a brief overview of a single hidden layer neural network model and one-class classification in Section \ref{sec:prelim}. This section also discusses the ELM framework and relevant online learning algorithms. Subsequently, we present the architecture and design details of the ADIC is followed by a discussion on ADEPOS scheme in Section \ref{sec:our_work}. Finally, section \ref{sec:results} contains the experimental results, on-chip characterization as well as health prediction results on the NASA bearing dataset\cite{NasaDataset} followed by the concluding remarks in the last section.

\section{Bakground: One-class classification, ELM and ADEPOS} 
\label{sec:prelim}
\begin{figure}[!ht]
\centering
\includegraphics[scale=0.3]{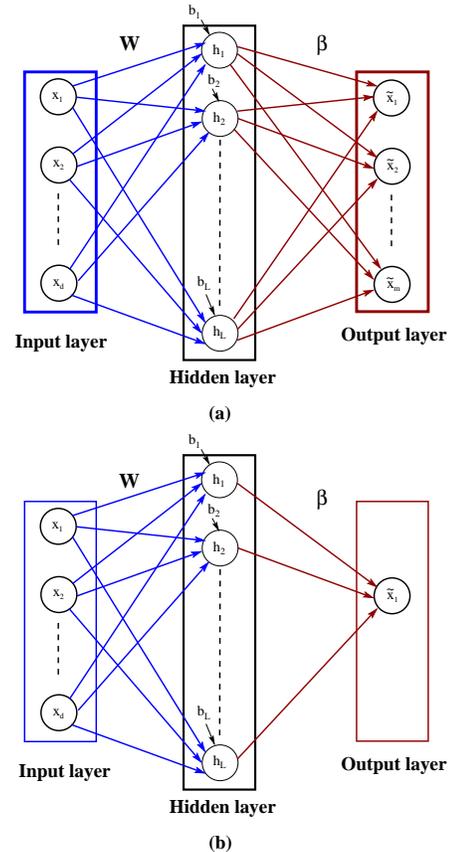}
\caption{A Single (Hidden) Layer Feed-forward Neural Network (SLFN): (a) input reconstruction to outputs and (b) output mapped to a single node \cite{Bose:ASPDAC2019, GAUTAM2017126}}
\label{fig1}
\end{figure}

A single layer feed-forward neural network, as depicted in Fig. \ref{fig1}(a), consists of input layer, $x = [x_1,\allowbreak x_2,\allowbreak \cdots,\allowbreak x_d]^T$, output layer, $\tilde{x} = [\tilde{x_1}, \tilde{x_2}, \cdots, \tilde{x_m}]^T$, and  hidden layer, $h = [h_1, h_2, \cdots, h_L]^T$. Here, $d$, $m$, and $L$ denote the respective number of input, output and hidden neurons. This network is most commonly referred to as an Auto Encoder (AE) engine when $m = d$. The AE network is trained to learn the input data distribution and reconstruct them at the output with minimal reconstruction error. While the hidden layer nodes encode an input vector $x$ into a feature space $h$ using the connection weights $W$ and biases $b = [b_1, b_2, \cdots, b_L]^T$, the output layer weights $\beta$ between the hidden and output neurons are used to decode the feature space embedded in $h$ aimed at reconstruction of the input features $x$. 

Details of the encoding and decoding process of AE are mathematically given in Eq. (\ref{eqn1a}) and (\ref{eqn1b}), where $g()$ refers to the neural activation function and $h_j$ denotes the output of the $j^{th}$ hidden neuron. 
\begin{equation}
h_j = g\left(\sum\limits_{i=1}^dW_{ji}x_i+b_j\right); ~~j=1,2,...,L
\label{eqn1a}
\end{equation}
\begin{equation}
\tilde{x_k} = \sum\limits_{j=1}^L\beta_{jk}h_j; ~~k=1,2,...,m
\label{eqn1b}
\end{equation}

Although the $sigmoid$ function is typically chosen as the activation function, $g(.)$, \textit{Rectified Linear Unit} ($ReLU$) has gained traction lately due to its effectiveness to solve the vanishing gradient problem and the ease of implementation on hardware.

\subsection{ELM based OCC}
We define a list of symbols in the following that will be used throughout this subsection:

\begin{tabular}{ll}
$W$     &input layer weights \\
$\beta$     &output layer weights \\
$X$     &input layer activation of N samples \\
$H$     &hidden layer activation of N samples\\
$x_i$     & input layer activation at $i^{th}$ input vector\\
$E_i$     &reconstruction error at $i^{th}$ input sample\\
$h_i$     &hidden layer activation at $i^{th}$ input sample \\
$\eta_i$     &learning rate at $i^{th}$ input sample \\
$K, \theta$  &intermediate variables for calculation\\
\\
\end{tabular}

A traditional autoencoder (TAE) requires a sufficiently large amount of data and computational resources in order to identify an optimal $W$ and $\beta$ values. The back-propagation (BP) method, although known to yield very accurate models, incurs high computational overhead due to its two-pass approach on each epoch. On the contrary, the training following the \textit{Extreme Learning Machine} (ELM) framework is faster where input weights, $W$, and biases, $b$ are generated randomly from a continuous probability distribution \cite{HUANG2006489}, \cite{NIPS2008_3495}. A \textit{Pseudo-random Binary Sequence Generator} (PRBS) for a given seed is able to generate a set of random numbers of specified data width in hardware implementation. This simplifies the network optimization for optimal output weights $\beta$.

In the batch learning approach \cite{HUANG2006489}, finding the optimal values of $\beta$ corresponding to least-square optimization requires to compute the hidden neuron outputs $H = [h_1,h_2, \cdots, h_N]$ for a batch of $N$ input data samples $\bar{X} = [x_1,x_2, \cdots, x_N]$. Since the output nodes (targets) are the same as training samples $\bar{X}$ for a reconstruction based AE, the optimal output weight $\beta^*$ is computed as shown below.
\begin{equation}
 \beta^*=H^\dagger \bar{X}
 \label{eqn2}     
\end{equation}

Notably, this equation reduces the task of parameter optimization for the entire AE network to a matrix inversion operation. Since these are non-square matrices, Moore-Penrose generalized inverse method \cite{PENROSE1954} is used to compute the pseudo-inverse $H^\dagger$ of $H$. 

Therefore, huge computational overhead, as well as large memory requirement, make this approach difficult to implement in hardware. In order to overcome these limitations, another version of ELM was proposed in \cite{Liang2006a}. Known as \textit{Online Sequential ELM} (OSELM), this method outlines the learning of output weight matrix $\beta$ based on a single data point $x_i$ in every iteration until convergence is attained. This approach is found to be similar to the \textit{stochastic gradient descent} method used in the back-propagation method since the output error due to a single input vector $x_i$ is used to update the output weight matrix $\beta^{(i-1)}$ to $\beta^{(i)}$. A set of equations describing this weight update process is presented in Eq. (\ref{eqn3a}) - (\ref{eqn3c}), details of which can be found in \cite{Liang2006a}.

\begin{equation}
K_i = K_{i-1} + H_{i}^T H_i
\label{eqn3a}
\end{equation}

\begin{equation}
E_i = x_i - H_{i} \beta_{i-1}
\label{eqn3b}
\end{equation}

\begin{equation}
\beta_{(i)} = \beta_{i-1} + K_{i}^{-1} H_{i}^T  E_{i}
\label{eqn3c}
\end{equation}

Nevertheless, the OSELM approach requires heavy computational overhead due to matrix inversion operation for each weight update as shown in Eq. (\ref{eqn3c}). These equations are valid for small batch of samples, say $N_i << N$, of $X_i = x_1,x_2, \cdots, x_{N_i}$., including $N_i = 1$ which refers to a single input sample $x_i$.

Recently, a simpler online weight update method was proposed in \cite{Tapson2013, VanSchaik2015a} which relies on the sequential update of the output weights using only one input sample $x_{i}$. This method is referred to as \textit{Online Pseudo-Inverse Update Method} (OPIUM). The set of equations that describe the OPIUM method are presented in Eq. (\ref{eqn3d}) - (\ref{eqn3f}). A close look at these equations reveal that (a) no matrix inversion is needed, and (b) weight update depends only on matrix multiplication. The matrix product $h_i^T\theta_{i-1}h_i$ in Eq. (\ref{eqn3d}) turns out to be a scalar entity, and hence eliminate the need for matrix inversion. Thus, all matrix multiplication operations can be realized in hardware using \textit{Multiply-and-Accumulate} (MAC) approach, which is a fundamental building block used in machine learning-based hardware designs.

\begin{equation}
\eta_i = \frac{\theta_{i-1} h_i}{1+h_{i}^T \theta_{i-1} h_i}
\label{eqn3d}
\end{equation}

\begin{equation}
\beta_{i} = \beta_{i-1}+\eta_i (x_i-\beta_{i-1}h_i)
\label{eqn3e}
\end{equation}

\begin{equation}
\theta_i = \theta_{i-1}-\eta_i \theta_{i-1} h_i
\label{eqn3f}
\end{equation}

This indicates that the usage of the ELM framework with the OPIUM learning algorithm is suitable for efficient hardware implementation of a single hidden layer feed-forward neural (SLFN) network. The combination of these methods reduces both computational overhead and energy consumption during training, without sacrificing on the quality of the trained model $\beta^*$ after convergence. The computational complexity of the OSELM method is estimated to be $\mathcal{O}(L^3)$, while that of OPIUM is $\mathcal{O}(L^2)$ for a network with $L$ hidden nodes. Moreover, OSELM versions require $3X$ higher memory space due to matrix inverse operation using the LU decomposition method.

To further reduce the computational complexity, the network in Fig \ref{fig1} (a) can be transformed to the one in Fig \ref{fig1} (b) having only one output node ($m = 1$) \cite{GAUTAM2017126}. We refer to the network modes in Fig \ref{fig1} (a) as \textit{reconstruction mode} and Fig \ref{fig1} (b) as \textit{boundary mode}. In this paper, we subsequently denote them as ELM-AE and ELM-B respectively. Unlike in ELM-AE, the single output node in ELM-B is set with any constant target value, e.g., $\tilde{x_1} = 1$, such that the output weight ($\beta^*$) is obtained after the training is expected to reconstruct a value of (or near to) $1$ at the output node \cite{GAUTAM2017126} for any $x_i$ that represents a healthy data. A faulty data $x_j$ exhibits a significant deviation from the reference value of $1$ depending on the quality of training of the output weight, i.e., the trained value $\beta^*$. The number of samples required for a good $\beta^*$ to converge depends on the network structure, i.e., whether ELM-AE or ELM-B. A comprehensive study on training convergence for different AE models has been summarized in \cite{Bose:ASPDAC2019}.

\subsection{Ensemble Learning}
\begin{figure}[!ht]
\includegraphics[scale=0.19]{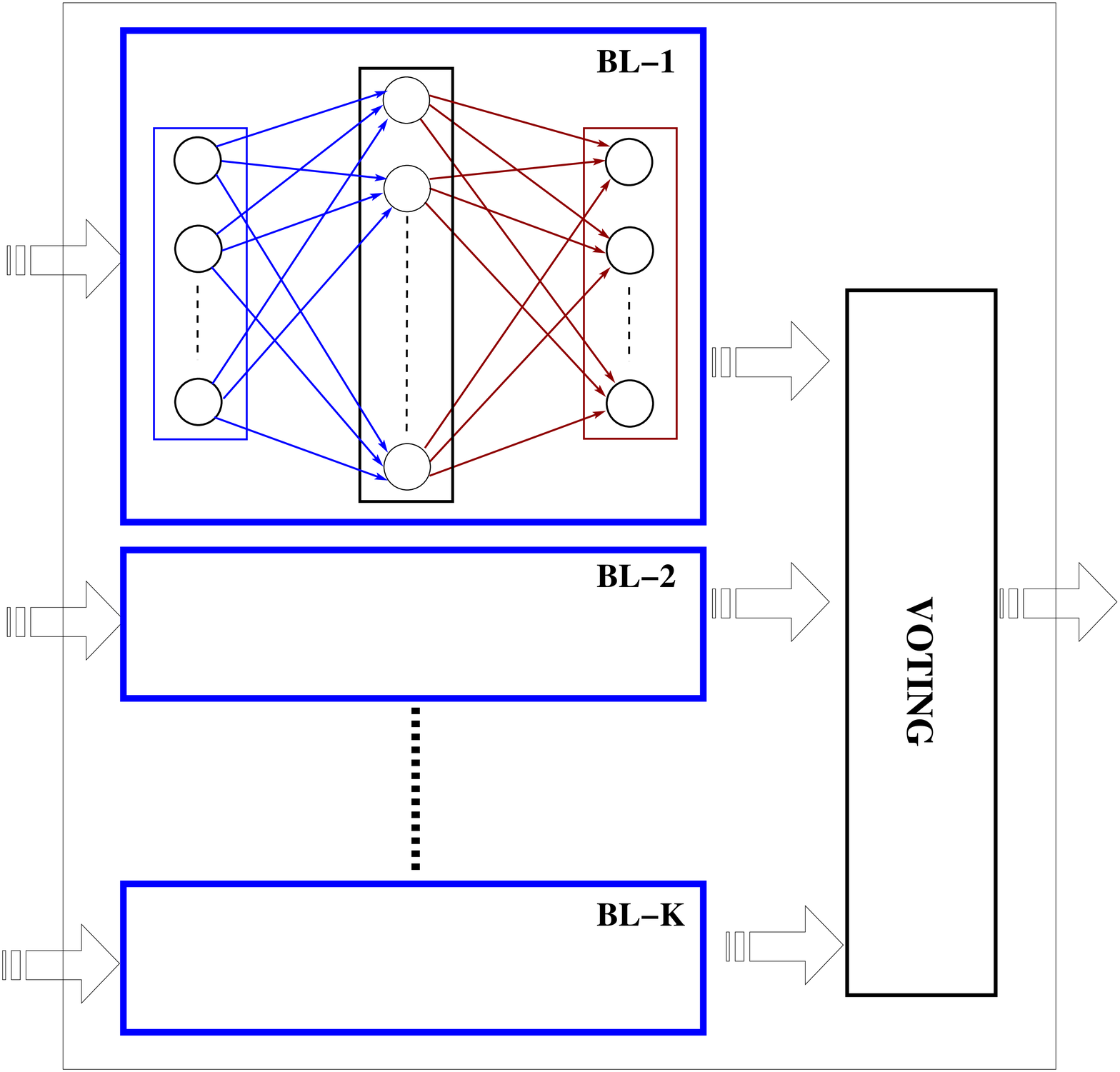}
\caption{An ensemble of $K$ OCC (BL) engines}
\label{fig_ensm}
\end{figure}

In Fig. \ref{fig_ensm}, we illustrate an ensemble of $K$ ELM based OCC engines, henceforth termed as a \textit{Base Learner} (BL) engine. Each BL engine is designed to be trained with the same set of input data $x_i$. However, the characteristic of each BL differs due to different seed values used in the PRBS module. This facilitates different distribution of input layer weights $W$ and biases $b$ and thus yields different trained $\beta$ values for each BL. Therefore, the performance of each BL during inference is expected to be different. This ensemble architecture can have a configurable number of active BLs (up to a maximum $K$ BLs) during inference while the rest remain inactive depending on the application. The output from each active BL engine can be used to generate the final output by several ensemble methods such as \textit{majority voting scheme}. In the case of majority voting, an odd number of \textit{active} BLs is required in order to avoid tied voting. Other than majority voting, this design can also support other ensemble applications such as \textit{AdaBoost} method \cite{8350948,ricardi14} in order to enhance system performance. However, this exercise is beyond the scope of this paper and considered to be addressed in our future works.

From the hardware implementation point of view, we can make any number of BLs inactive (hibernate) once their training is done by simply putting their respective clocks off. This strategy has a direct impact on dynamic energy-saving and is described in the subsequent section. Dynamic voltage scaling (DVS) is also feasible for this architecture, by scaling down the supply voltage $V_{dd}$ of the respective inactive BLs to a permissible value, without corrupting the trained model parameters due to memory failure at lower supply voltages. This will further help in power consumption by reducing the leakage power of both the active memories and the standard logic in those inactive BLs.

\subsection{The ADEPOS algorithm}
Now, we briefly discuss the ADEPOS algorithm \cite{Bose:ASPDAC2019} for energy-saving suitable for the proposed OCC engine ADIC. During early life, a machine presumably remains to be healthy and unlikely to show any fault due to wear and tear. During this healthy period, the output of the OCC-engine will not flag error indicating any failure in the machine. Thus the detection of an anomaly at this stage is rare unless caused by unexpected failures of the machine or even spurious signals coming from the sensor module. This fact is utilized by the ADEPOS algorithm which is based on the hypothesis that we can employ inaccurate computing at this stage of early inference. Errors due to this approximation can be overcome by careful inspection of the fault and its trend of crossing the threshold. The pseudocode in Algorithm \ref{alg1} outlines the ADEPOS algorithm, showing only the variation in the number of active BLs. 

\begin{algorithm}
\caption{ADEPOS Algorithm}
\label{alg1}
\begin{algorithmic}
\REQUIRE Maximum number of BL: $K$; Threshold of $i^{th}$ BL: $Th_i$; Input feature vector at instant t, $S_t$; Model of $i^{th}$ BL: $f_i(W,b)$;
\ENSURE Ensemble output, $O_{e}$;
\STATE $O_e \leftarrow 0;~N\leftarrow 1$; 
\WHILE{$O_e \neq 1$}
\STATE {$O_{e,l}\leftarrow 1$};
\WHILE{$O_{e,l} \neq 0$}
\STATE {$~T\leftarrow 0$};
\FOR{$i=1;~i\leq N;~i=i+1$}
\STATE {$O_i\leftarrow f_i(S_t,W,b)$};
\STATE {$D_i\leftarrow \sum_{features} (O_i-S_t)^2 > Th_i$};
\STATE {$T\leftarrow T+D_i$};
\ENDFOR
\IF{$T < \frac{N+1}{2}$}
\STATE {$O_{e,l}\leftarrow 0$};
\STATE {$O_e\leftarrow 0$};
\IF{$N\neq 1$}
\STATE {$N\leftarrow N-2$};
\ENDIF
\ENDIF
\IF{$T\geq \frac{N+1}{2}$}
\STATE $O_{e,l} \leftarrow 1$;
\IF{$N\neq K$}
\STATE {$N\leftarrow N+2$};
\ELSE
\STATE $O_e \leftarrow 1$;
\ENDIF
\ENDIF
\ENDWHILE
\STATE {$S_t \leftarrow S_{t+1}$};
\ENDWHILE
\STATE{Anomaly Detected}
\end{algorithmic}
\end{algorithm}

The effective number of hidden neurons in the proposed ensemble framework, $L_k$, is given by:
\begin{equation} 
L_k = K\times L
\label{eq:Leff}
\end{equation}
where $K$ is the number of base learners (BL) in the ensemble network. 

From the hardware viewpoint, the training of a larger neural network of $K\times L$ hidden neurons requires a large memory of the size of $K^2\times L^2$ for storing $\theta$ (Eq. (\ref{eqn3f})), whereas an ensemble of $K$ BLs with $L$ neurons require only $K\times L^2$ memory. This saves memory area by a factor of $K$. 


\section{ADIC Hardware Architecture} 
\label{sec:our_work}
\begin{figure}[!ht]
\centering
\includegraphics[scale=0.26]{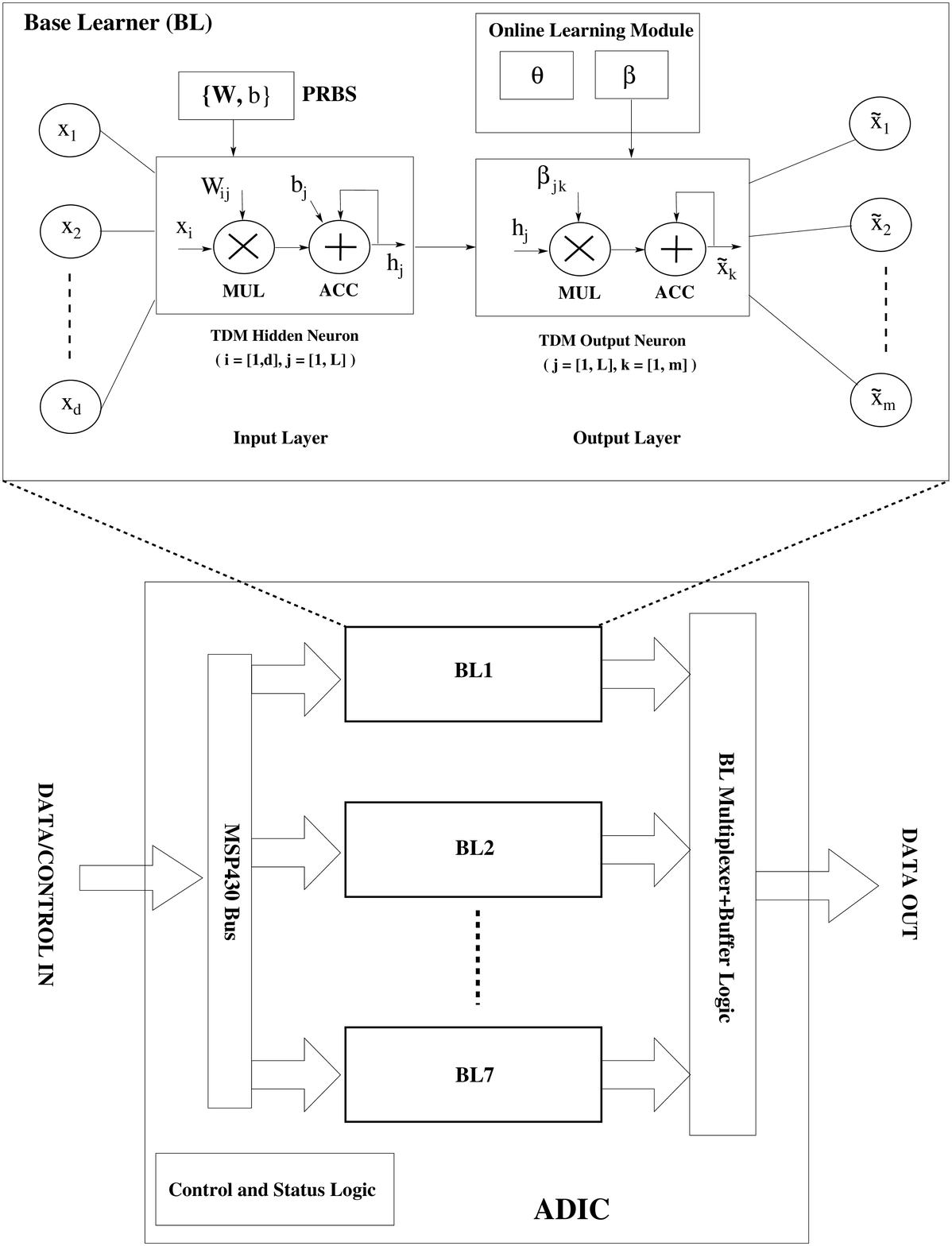}
\caption{Architecture of ADIC and one Base Learner (BL) module}
\label{fig_arch}
\end{figure}

Here, we discuss the proposed hardware architecture of ADIC and its implementation that supports online learning (via OPIUM) and approximate computing (via ADEPOS). Fig. \ref{fig_arch} depicts an overview of the architecture for the proposed ELM based OCC engine, alternatively called BL engine in the ensemble discussed in the previous section. Each BL consists of a PRBS module, a TDM (time division multiplexing) hidden and output neuron, and an online learning module. As discussed in the previous section, the training of $\beta$ values in each BL is done with different seed values for the corresponding PRBS module in it. This pseudo-random generator module generates a maximum $16{-}bit$ random number for the input weights $W = \{w_{ij} | i \in \{1, d\},  and,  j \in \{1, L\} \}$ and biases $b = \{b_j | j \in \{1, L\} \}$ within the dynamic range of $-2^{15}$ to $2^{15} - 1$. The implementation of the PRBS module eliminates the need to store the input weights and biases and the need to obtain their trained values as in the case of back-propagation based methods. Thus, the memory requirement to hold the trained parameters for this network is reduced by $\approx 50\%$. Additionally, this approach reduces the training overhead for $\{W, b\}$. Power analysis using a post-layout extracted netlist in INNOVUS shows that the combined logic of PRBS and MAC in hidden layer computation consumes $3.5X$ less power over the MAC and the memory cells used to store the input weights $W$ and biases $b$. Moreover, we achieve an area reduction of $80\%$ in the hidden layer by implementing PRBS. This eliminates the need to store the random weights and biases $\{W, b\}$ for the first stage of ELM based SLFN network only.

\begin{figure}[!ht]
\centering
\includegraphics[scale=0.425]{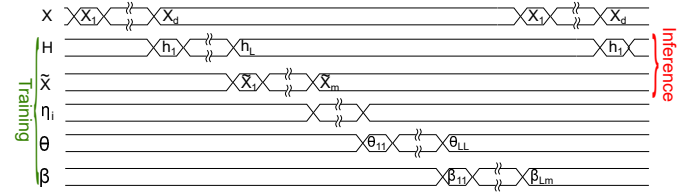}
\caption{Timing diagram of the dataflow of on-chip calculation during the training and testing of a BL.}
\label{timingDiagram}
\vspace{-1em}
\end{figure}

The proposed ADIC implements the online learning framework OPIUM \cite{Tapson2013} in order to reduce the computational overhead per input sample, $x = \{x_i | i \in \{1:d\}\}$, yielding $3X$ reduction in memory requirement over OSELM method. This reduces the total power consumption due to the inverse calculation required in the OSELM method. In this design, we adopted a single physical neuron that works in \textit{time-division-multiplexing} (TDM) fashion for both the $1^{st}$ and $2^{nd}$ layers of the ELM as depicted in Fig. \ref{fig_arch}. This approach emulates the scenario of $L$ physical hidden nodes and also $m$ output neurons, by suitable configuration of $L$ and $m$. This approach not only reduces the requirement for large area requirement for logic and memory but also reduces the power consumption due to the number of computes depending on the value of these parameters. In the design, $L$ can vary in the range of $1-32$ while $m$ can vary in the range of $1-d$ with a maximum value of $d = 16$. As a result, this architecture also provides the required flexibility to configure the ADIC as a boundary based ELM engine for one class classification by programming the number of output neurons $m = 1$ and $\tilde{x_1} = 1$. This enables trading off the higher processing time required for reconstruction mode with the quality of inference in boundary mode. As mentioned earlier, the number of output neurons $m$ for reconstruction mode remains the same as the number of input neurons $d$. With these parameters, including the number of hidden nodes $L$, the entire network can be made configurable. These configurable parameters also dictate the PRBS module to accordingly generate the required number of input weights $W$ and biases $b$, having the respective dimensions of $L\times d$ and $L \times 1$. 

This design has configurable datapath ranging from $8$ to $16$ bits in steps of $4$ bits, while the number of input ($d$) and output neurons ($m$) can also be varied from $1$ to $16$. Moreover, the number of hidden nodes $L$ is made configurable and go up to $32$ in each base learner (BL) module. In order to reduce the area and the leakage power due to SRAM cells, we restrict the maximum datapath width to $16bit$ (signed integer) for encoding $W$, $b$, $\beta$ as well as the input data $x_i$. Additionally, we provide another level of bit precision control by independently controlling the bit width of the random numbers generated by the PRBS module for $W$ and $b$, and can range from $2$ bits to a maximum of $8$ bits in steps of $2$ bits. It is important to note that these bit-precision control approaches apply to inference mode only, while the training is always done with the highest allowable precision of $16$ bits. This ensures that the training phase gets the highest accuracy of computations. 

In addition to that, a method to reduce power consumption during training by reducing computational overhead during training was proposed in \cite{VanSchaik2015a} as a \textit{Lite} version of OPIUM. This method is commonly referred to as \textit{OPIUM-Lite}. As discussed in \cite{VanSchaik2015a}, we introduce a provision to prohibit the update of $\theta$ variable and initialize it with a configurable value, in order to realize the OPIUM-Lite version. The corresponding SRAM module in each BL that stores $\theta$ values and the logic around it are completely turned off by clock-gating. This technique, by prohibiting the read/write access, reduces the dynamic power significantly as it is the largest memory cell used and has almost double the combined size of all other SRAM cells used in each BL. 

 Figure \ref{timingDiagram} presents a timing diagram of the dataflow of on-chip calculation during the training and testing of a BL. ADIC follows Eq. (\ref{eqn1a})-(\ref{eqn1b}) for hidden and output neurons calculation and Eq. (\ref{eqn3d})-(\ref{eqn3f}) for parameter updates during online training. For illustration, we show the update of all the elements ($h_j/\tilde{x_m}/\beta_{jk}$) of each parameter in every clock cycle. However, the calculation of each element takes several clock cycles except the input (X) dataflow. Even though all the parameters are $16$-bit precision, calculations of the partial sum are done in higher bit precision to avoid overflow. For instance, we use $32$-bit for the calculation of ACC in hidden and output neurons (Fig. \ref{fig_arch}). However, based on the 2-bit configuration, we choose intermediate 16-bit of final ACC calculation, such as ACC[27:12] for hidden and output neurons ($H$ and $\tilde{X}$).
 
\begin{figure}[!ht]
\centering
\begin{subfigure}[b]{0.23\textwidth}
\centering
\includegraphics[scale=0.3]{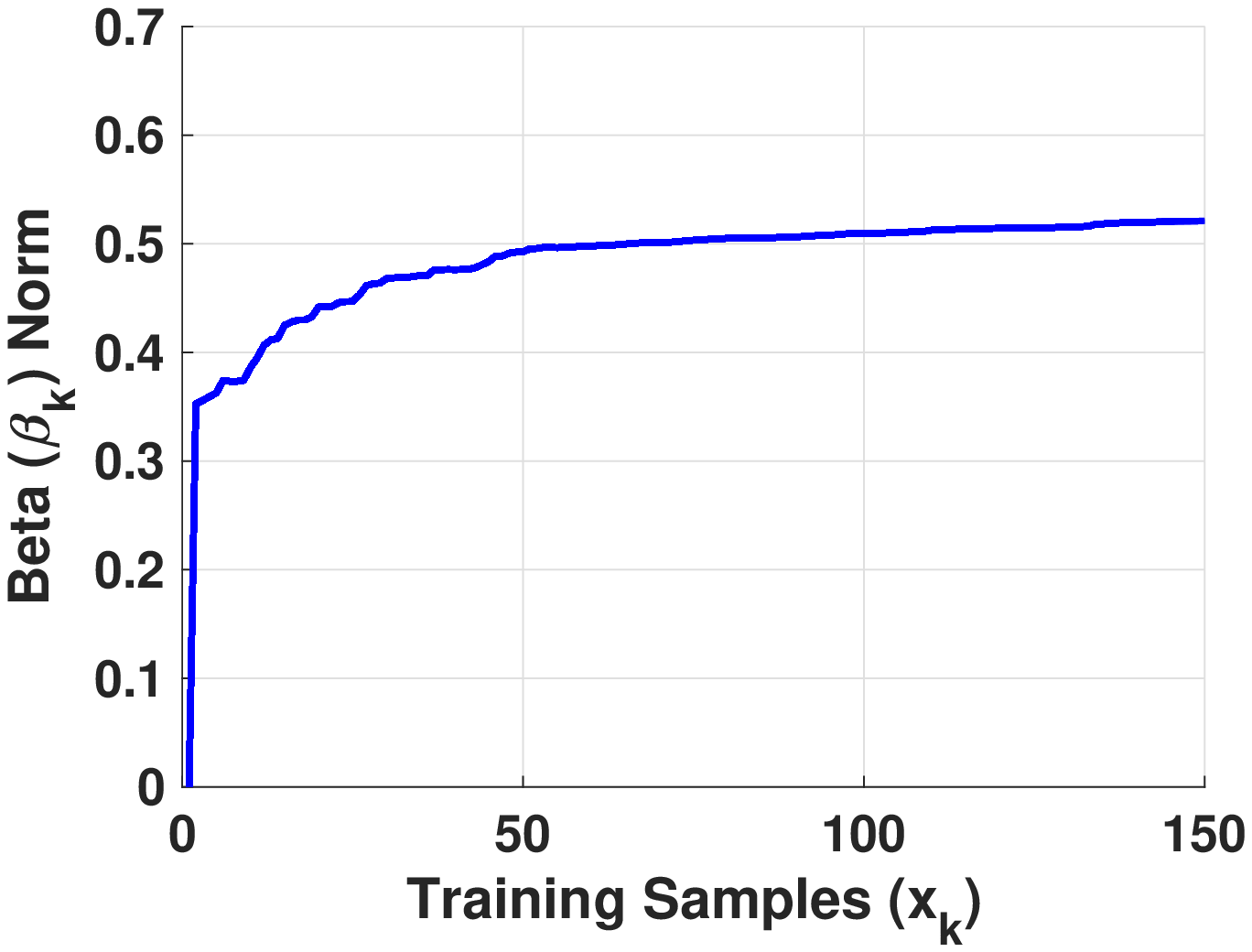}
\caption{}
\end{subfigure}
\begin{subfigure}[b]{0.23\textwidth}
\centering
\includegraphics[scale=0.3]{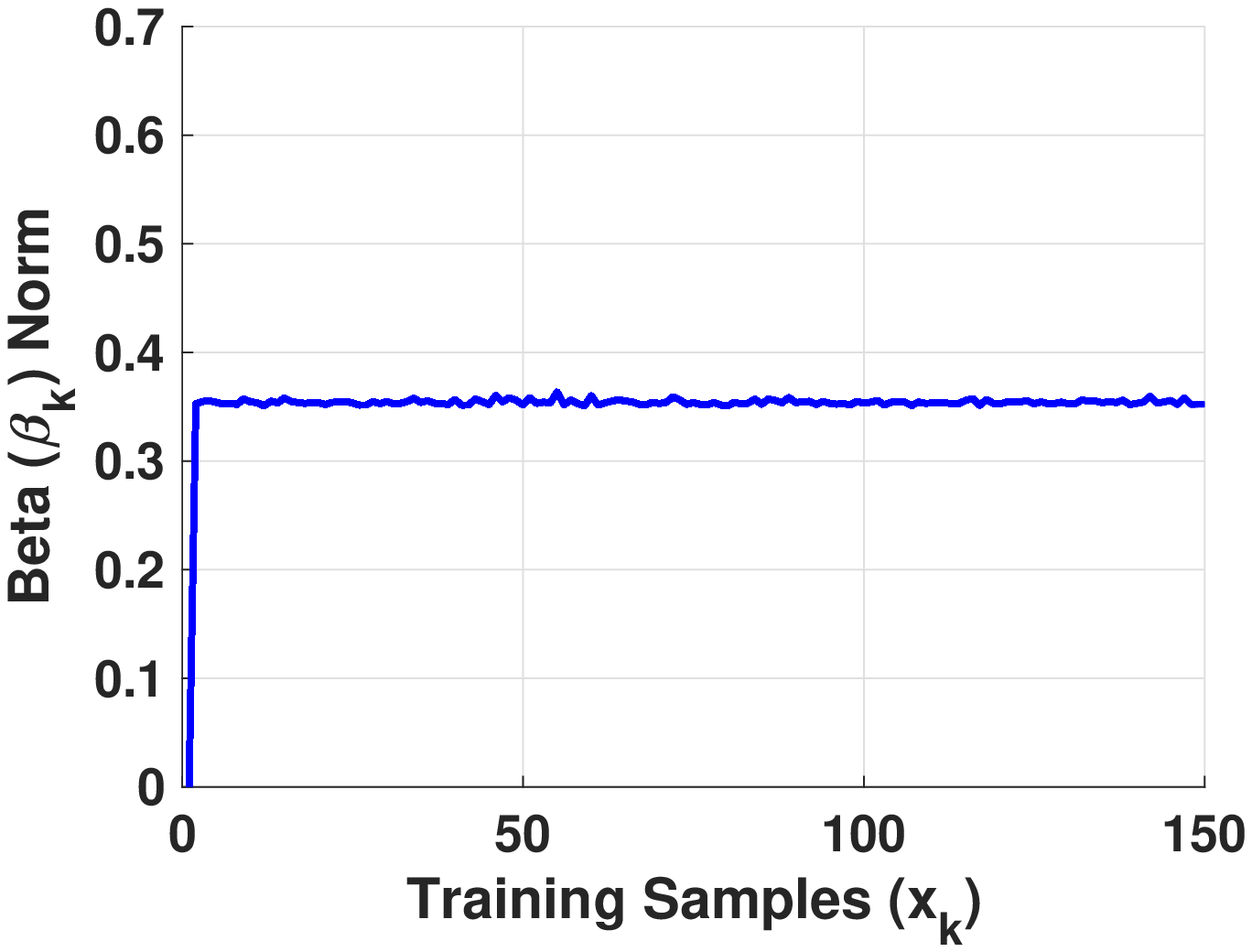}
\caption{}
\end{subfigure}
\begin{subfigure}[b]{0.23\textwidth}
\centering
\includegraphics[scale=0.3]{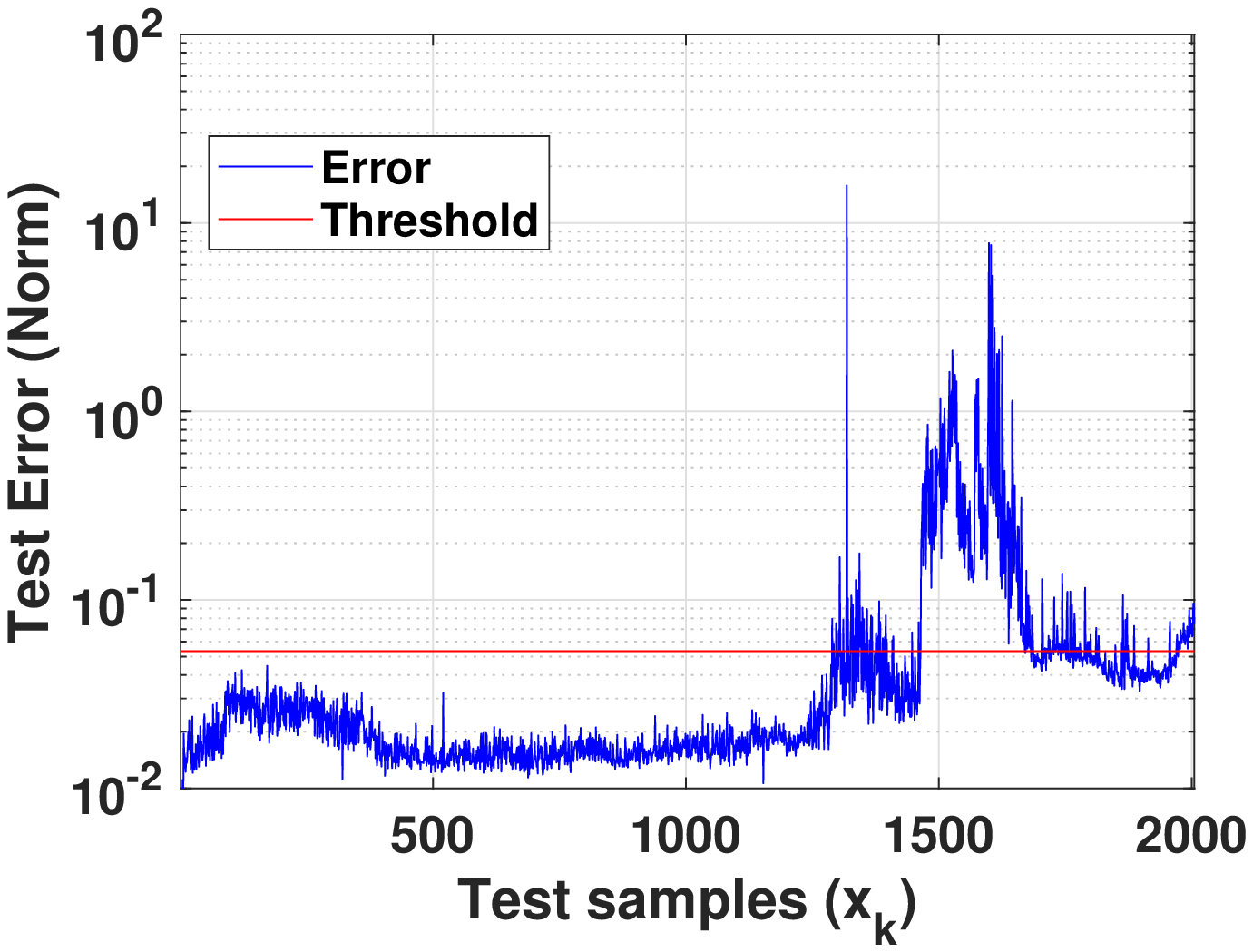}
\caption{}
\end{subfigure}
\begin{subfigure}[b]{0.23\textwidth}
\centering
\includegraphics[scale=0.3]{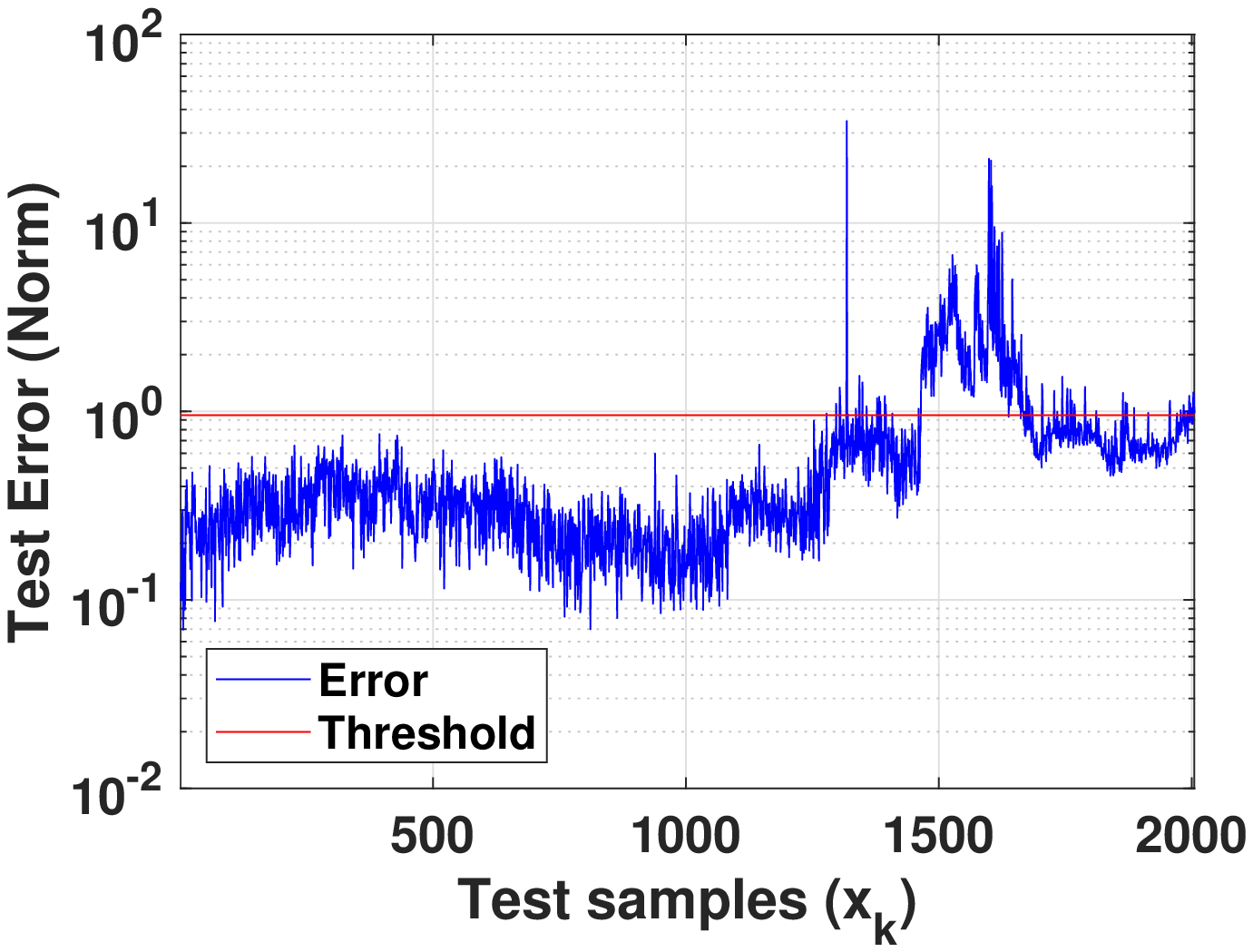}
\caption{}
\end{subfigure}
\caption{Matlab simulation for Time-Error Response of a faulty machine in NASA bearing dataset \cite{NasaDataset}: (a) OPIUM training, (b) OPIUM-Lite training, (c) OPIUM inference, and (d) OPIUM-Lite inference}
\label{fig_err_matlab}
\vspace{-1.5em}
\end{figure}

In Fig. \ref{fig_err_matlab}, we present Matlab simulation results for commonly used machine health monitoring dataset \cite{NasaDataset}, using both OPIUM and OPIUM-Lite training for the same number of samples. This dataset exhibits a healthy initial part and therefore is used for training. While most of the remaining data points represent a healthy state, data points beyond $1600$ tend to show increasing fault levels due to higher vibration levels. The results show that OPIUM-Lite takes relatively less number of input samples ($x_i$) to converge than OPIUM method, as depicted in Fig. \ref{fig_err_matlab} (a) and (b) for OPIUM and OPIUM-Lite respectively. Notably, the norm of $\beta$ appears to be noisier in the case of OPIUM-Lite than OPIUM even towards the end of the training. Likewise, the inference response of OPIUM-Lite based training (\ref{fig_err_matlab}(d)) appears to be noisier than OPIUM method (\ref{fig_err_matlab}(c)), specially during the period of good health.  However, this does not adversely affect the performance due to two reasons: first, the threshold in case of OPIUM-Lite is proportionately increased (red line in \ref{fig_err_matlab}(c) \& (d)) and second, the relative change in test error on encountering an actual anomaly is still much larger than the baseline and crosses the threshold at the same sample index.
 We calculate the test error as $\norm{\mathbf{x}-\tilde{\mathbf{x}}}$, where $\textbf{x}$ and $\tilde{\textbf{x}}$ denote the input and reconstructed output vectors respectively. In ADIC, there is a provision of choosing between OPIUM and OPIUM-Lite methods depending on the application and thus exploits the energy-saving capabilities of OPIUM-Lite due to the lesser number of samples used for training, without any significant compromise on fault detection. The prohibition to update $\theta$ in OPIUM-Lite also incurs significant energy savings. Relevant experimental results are presented in Section \ref{sec:results}. 

In the current implementation of ADIC, we dedicate different power ($V_{dd}$) lines for each BL so that their leakage power can also be controlled while they are inactive during inference. Since the learning module is independent of the inference logic in each BL, we can completely shut off the power supply $V_{dd}$ to all the SRAM cells used for training only. For example, the memory used for storing $\theta$ parameters within the respective BLs can be completely turned off once the learning of $\beta$ reaches a point of convergence. This aims to reduce the leakage power drawn by these training memories. The availability of \textit{isolation cells} available in TSMC $65nm$ Low power (LP) process library helps maintain the functioning of the rest of the logic. In this design, the maximum number of active BLs is fixed to $7$ due to the targeted die area constraint of $2mmX2mm$. 

\section{Results}
\label{sec:results}
In this section, we first present results of the characterization of the ADIC in terms of operational power supply range, clock frequency and power dissipation. Later, we present results of applying ADIC to a predictive maintenance task to determine health condition of bearings.
\begin{figure}[!ht]
\centering
\begin{subfigure}[b]{0.24\textwidth}
\includegraphics[width=1.7in,height=1.5in]{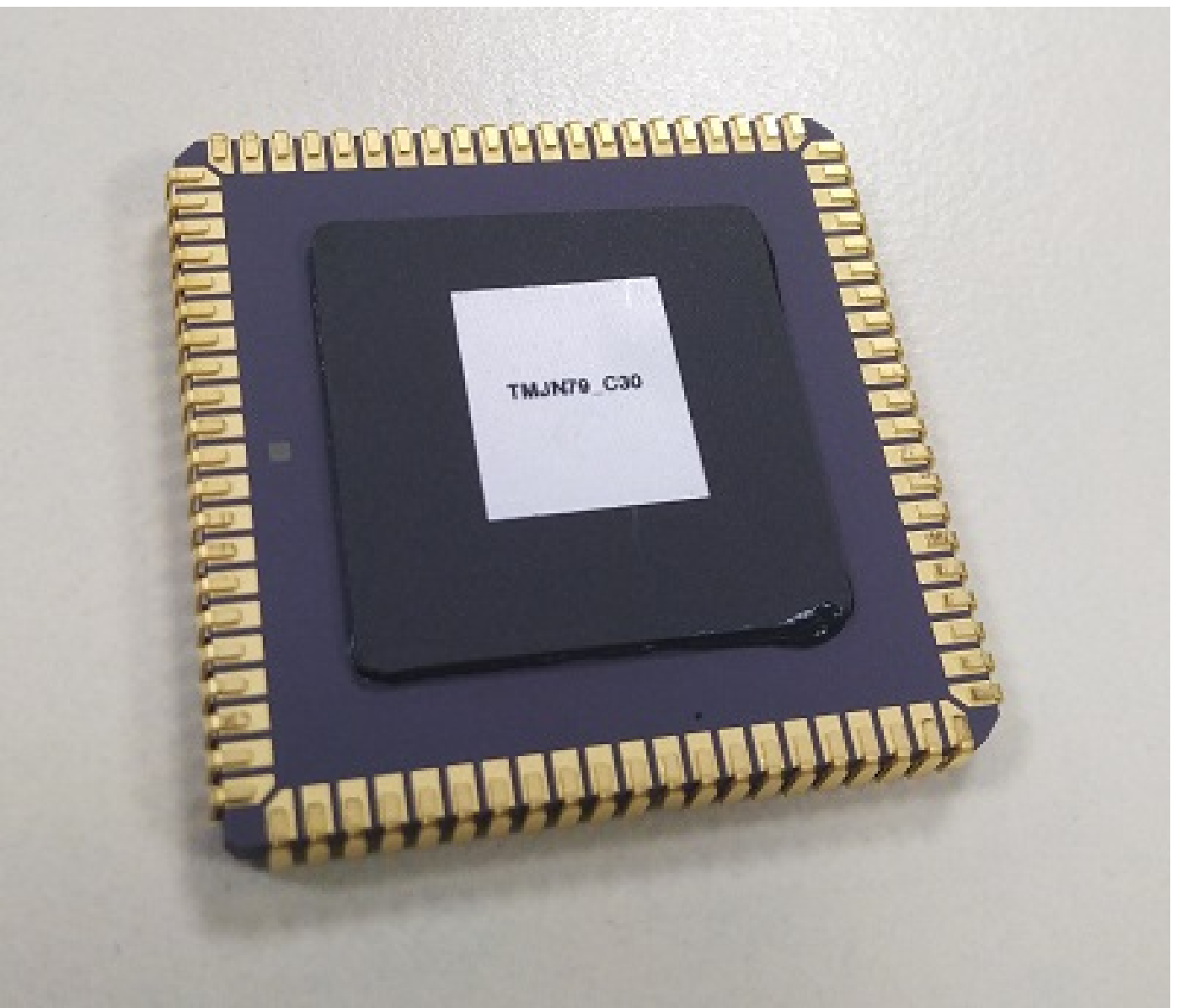}
\caption{}
\end{subfigure}
\begin{subfigure}[b]{0.24\textwidth}
\includegraphics[width=1.6in,height=1.5in]{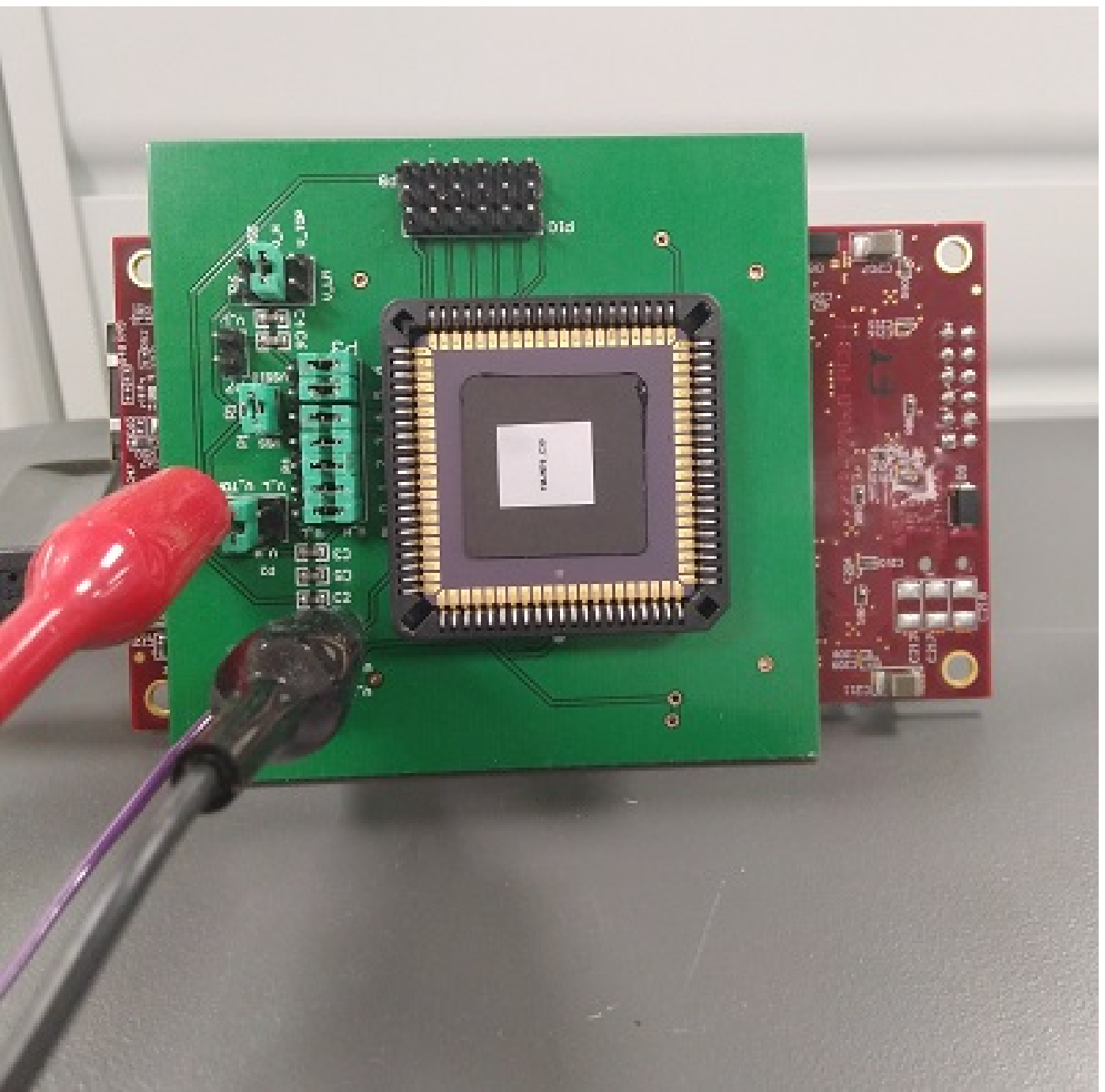}
\caption{}
\end{subfigure}
\caption{Experimental Setup: (a) fabricated device \textit{ADIC}, and (b) the test infrastructure}
\label{fig_tchip_setup}
\vspace{-1.5em}
\end{figure}

\subsection{Characterization of \textit{ADIC}}
Figure \ref{fig_tchip_setup} shows the packaged chip for \textit{ADIC}, and the test setup for online learning and machine health monitoring. Even though in this setup, an FPGA board sends the extracted features to \textit{ADIC} for ease of testing, other low power microcontrollers can be employed to transfer the data as well to reduce system power in deployed solutions. We vary the $Vdd$ of \textit{ADIC} from $0.75V$ to $1.2V$, while the input clock frequency is varied from $10MHz$ to $50MHz$. The current implementation of \textit{ADIC} supports $16$-bit data, as well as up to $14$-bit address bus and the control signals for read/write operations. In the next two subsections we will discuss the characterization of \textit{ADIC} with respect to supply voltage, frequency and number of active BLs.

\subsubsection{Dynamic Voltage and Frequency Scaling (DVFS)}
\begin{figure}[!ht]
\centering
\begin{subfigure}[b]{0.24\textwidth}
\centering
\includegraphics[scale=0.32]{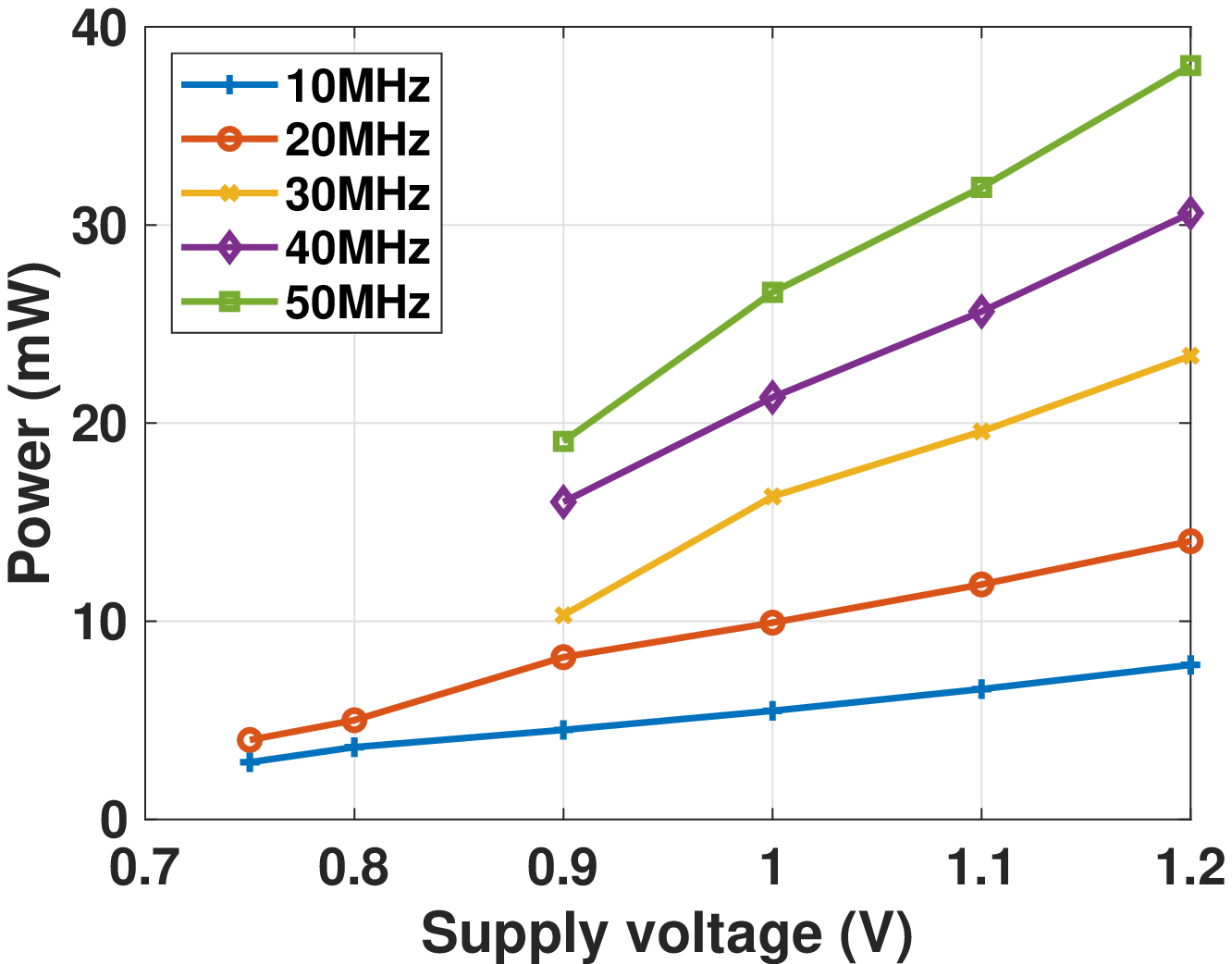}
\caption{}
\end{subfigure}
\vspace{0.1in}
\begin{subfigure}[b]{0.24\textwidth}
\centering
\includegraphics[scale=0.32]{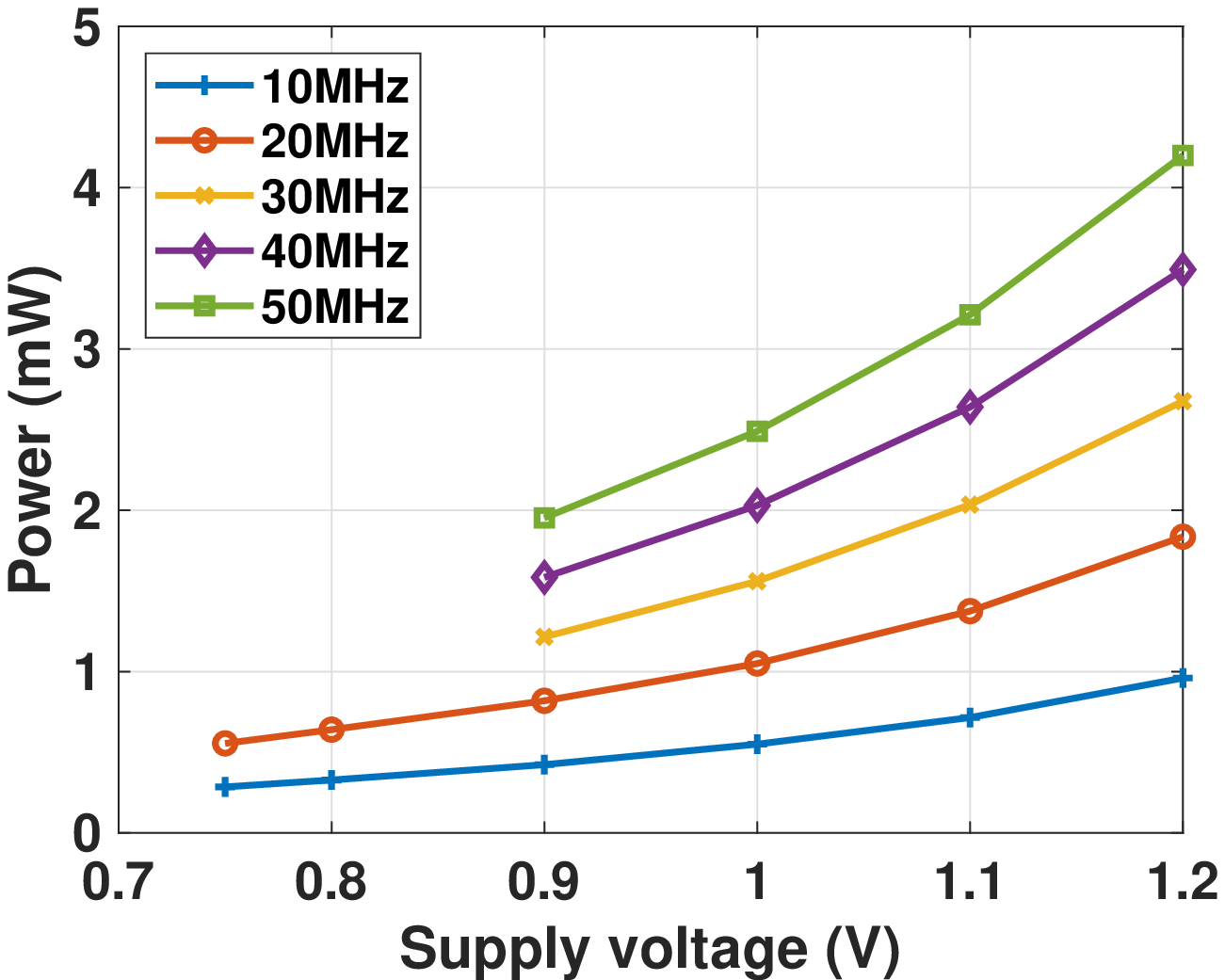}
\caption{}
\end{subfigure}
\caption{Power ($mW$) vs Supply voltage $V_{dd}$ ($V$) at different clock frequencies ($10$-$50MHz$) and $K = 7$: (a) Training, and (b) Inference.}
\label{fig_vdd_freq}
\vspace{-1.5em}
\end{figure}

We present the power consumption of \textit{ADIC} across different supply voltages as well as frequencies in Fig. \ref{fig_vdd_freq}(a)-(b). These results are presented for both training and inference phases of \textit{ADIC} for the maximum number of active BLs ($K$=$7$) in the system. It can be seen from the figures that the device is functional to a minimum $V_{dd}$ of $0.75V$ although $1.2V$ is the nominal supply voltage. For both training and inference mode, the trend clearly shows that we have power saving of (a) $3X$ when $V_{dd}$ is reduced from $1.2V$ to $0.75V$ at $20MHz$, and (b) $2.5X$ when $V_{dd}$ is reduced from $1.2V$ to $0.75V$ at $10MHz$. Also, it can be seen that the minimum required supply voltage for clock frequencies above $40$MHz turns out to be $V_{dd}$ = $0.9V$.

\begin{figure}[!ht]
\centering
\includegraphics[scale=0.55]{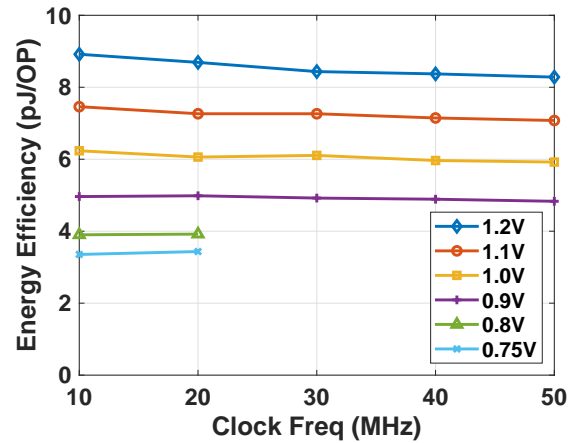}
\caption{Inference Energy Efficiency of \textit{ADIC} at different $V_{dd}$ and frequencies.}
\label{fig_vdd_freq2}
\end{figure}

The previous plots show reduction of both power and throughput by DVFS. Next, figure \ref{fig_vdd_freq2} presents the inference energy per operation of \textit{ADIC} for the DVFS data presented earlier. \textit{ADIC} takes $\approx 1000$ clock cycles to generate an output decision for a single BL in the inference phase. This includes the cycles needed to transfer the data from the input layer to the output layer.  Evidently, a trade-off exists between energy consumption and the throughput of the system. \textit{ADIC} can achieve a maximum of $10K$ classifications per second at the cost of $3.35 pJ/OP$ at $10MHz$ and $0.75$V. On the contrary, for the \textit{ADIC} to cater a higher throughput of say $50K$ classifications per second, the clock frequency is required to be increased to $50MHz$ leading to the corresponding increase in energy from $4.83 pJ/OP$ at $0.9V$ to $8.28 pJ/OP$ at $1.2$V.  As an application (PdM) point of view, we can operate the system at the lowest frequency possible due to the lower data rate. However, a faster classification rate is preferable since this also implies lesser time to classify one sample and more time to be spent in low-energy sleep mode.

Hence, depending on the applications that demand higher classification throughput, the ADIC can be run at the most favorable $V_{dd}$ and clock frequency for the best energy per operation value. It is noticeable from the plots in Fig. \ref{fig_vdd_freq2} that the energy efficiency of \textit{ADIC} decreases with higher $V_{dd}$. However, these variations are practically insignificant with respect to the variation in clock frequency running at the same $V_{dd}$ value. This is due to the fact the dynamic power consumption ($\propto CV_{dd}^2f_{clk}$) in \textit{ADIC} is the most dominant component in total power consumption while the static power (which gets amortized with higher clock frequency) is negligible. For example, these plots show a marginal variation in $pJ/OP$ with respect to clock frequency at $V_{dd} = 1.2V$ and get flatter as the $V_{dd}$ is reduced to a lower working value. This is attributed to the further decrease in static power with the corresponding decrease in $V_{dd}$.

As it is also depicted in Fig. \ref{fig_vdd_freq}, the minimum supply voltage $V_{dd}$ required to run at a given clock frequency varies in order to ensure that the timing of the most critical paths in the design are not violated. For example, $V_{dd} = 0.75V$ can handle an operating frequency up to $20MHz$, while $V_{dd} = 0.9V$ is required for running the design for a higher clock frequency of $30-50 MHz$. 

\subsubsection{Power consumption Vs \# active BLs}
\begin{figure}[!ht]
\centering
\includegraphics[scale=0.36]{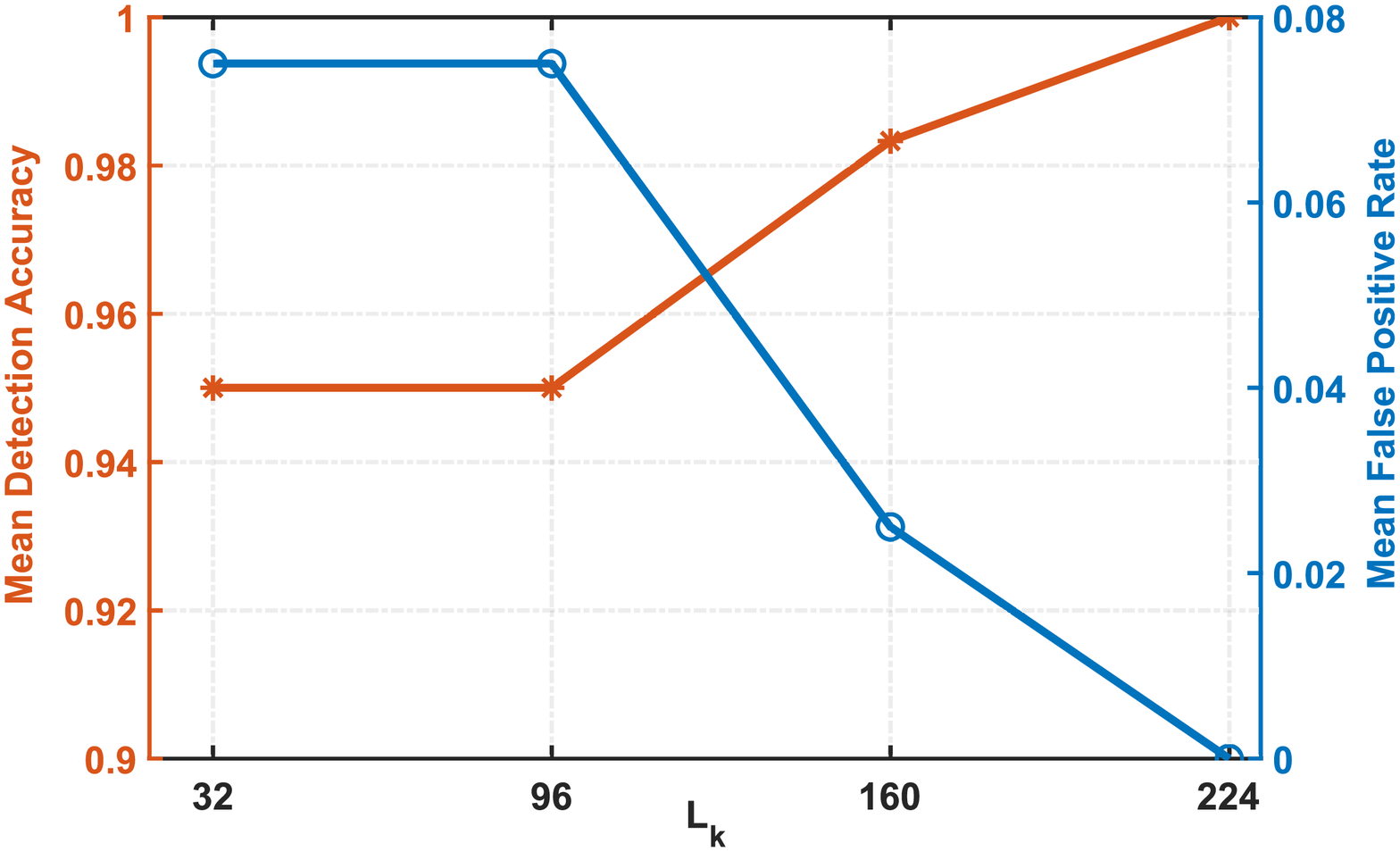}
\caption{ Mean detection accuracy and false positive rate vs. effective number
of hidden neurons ($L_k= K\times L$) in the network. $K \in \{1,3,5,7\}$ and $L=32$.}
\label{accuracy_curve}
\end{figure}

 Figure \ref{accuracy_curve} presents the mean detection accuracy and false positive rate of  ADIC across a different number of hidden neurons in the network using the NASA bearing dataset.
  As long as the accuracy of each BL (weak learner) is higher than the random guess ($50\%$), we can build a stronger model by an ensemble of  multiple weak models. Therefore, even though the mean detection accuracy of each BL is $95\%$ due to its simple architecture, we achieved $100\%$ detection accuracy by an ensemble of $7$ BLs. Moreover, since we are getting $100\%$ detection accuracy deploying the single hidden layer network in conjunction with the ADEPOS algorithm, we did not increase the depth of our architecture beyond two layers.
 
\begin{figure}[!ht]
\centering
\begin{subfigure}[b]{0.24\textwidth}
\centering
\includegraphics[scale=0.32]{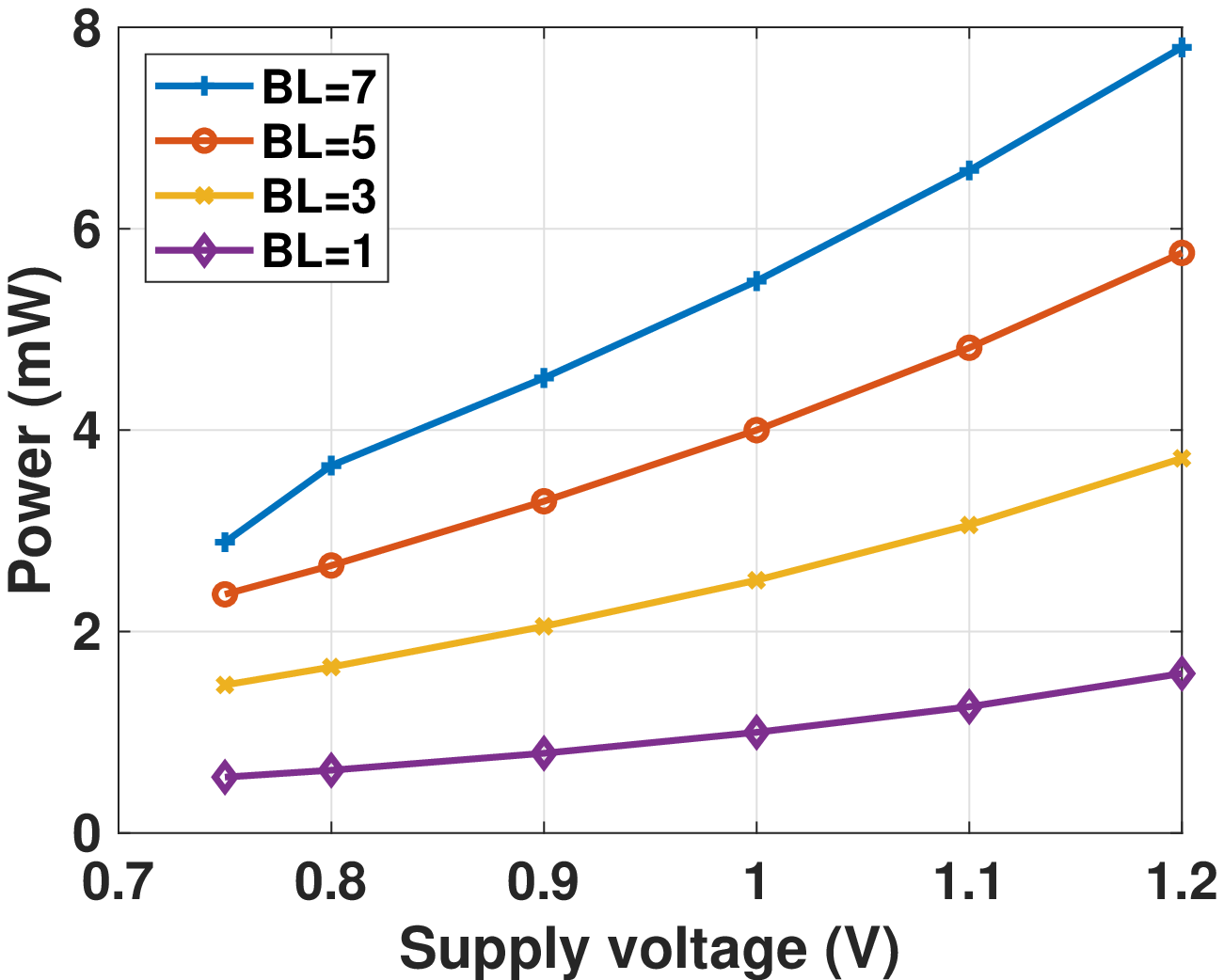}
\caption{}
\end{subfigure}
\vspace{0.1in}
\begin{subfigure}[b]{0.24\textwidth}
\centering
\includegraphics[scale=0.32]{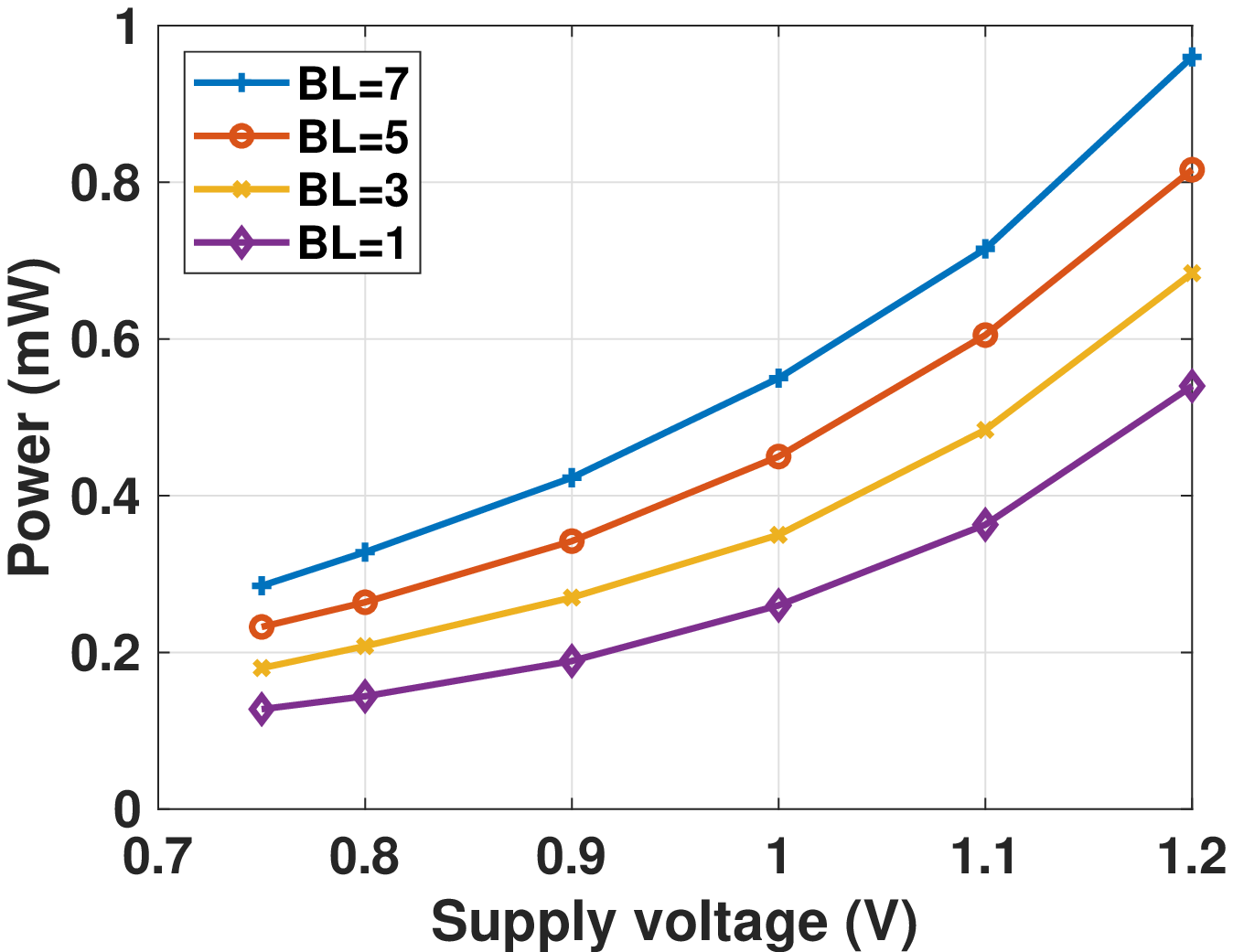}
\caption{}
\end{subfigure}
\caption{Power consumption of \textit{ADIC} vs Supply voltage $V_{dd}$ ($V$) for different number of active BLs ($K \in \{1,3,5,7\}$) at $10MHz$: (a) Training, and (b) Inference.}
\label{fig_vdd_bl}
\end{figure}

As discussed in Section \ref{sec:our_work}, the existing ADEPOS algorithm achieves energy savings by varying the number of active BLs ($K$) during the inference phase. The results in Fig. \ref{fig_vdd_bl} (b) show how the variation in $K$ can be beneficial in reducing the power consumption during the inference phase when the machine is in good condition and the classification can be done with majority voting among a fewer number of active BLs. During the early life of a machine, $K$ can be as low as $1$ yet giving acceptable results showing good health of the machine. However, as the ADEPOS algorithm dictates, on detection of any error by the current active BLs, $K$ can be increased to monitor the recurrence of the error on the same data. This process continues while increasing the current $K$ to the next allowable value if the error persists. If $K$ reaches the maximum value of $7$ (supported in this implementation) and error is detected on the same input data sample from the sensors, then the machine is identified as a faulty one. It can be seen that a $1.8X$ reduction in power consumption is possible when running with $K = 1$ compared to $K = 7$ at $V_{dd}$ = $1.2$V. Moreover, this algorithmic power reduction is invariant of supply and operating frequency of \textit{ADIC}.

 Since the ADEPOS algorithm demands that all the BLs should be trained so that they are ready for the inference phase, we also plot the power requirement of a different number of BLs during the training phase in Fig. \ref{fig_vdd_bl}(a). In general, we expect that the energy consumption overhead for training is much less as compared to the inference energy. This is due to the fact that the training is done with a relatively smaller (limited) number of input samples at the beginning of deployment, while the inference can go as long as the energy source to the device is not depleted or until the fault is detected in the machine.  

\subsection{Energy Saving in Inference: ADEPOS}
\begin{figure}[!ht]
\centering
\includegraphics[scale=0.36]{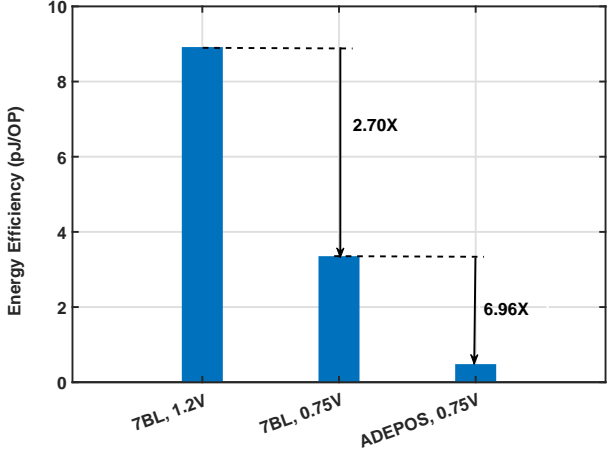}
\caption{Comparing Energy Efficiency ($pJ/OP$) for Inference: \textit{ADIC} at \{$1.2V, 0.75V$\} vs \textit{ADIC}+ADEPOS at {$0.75V$}}
\label{fig_adepos_cmp}
\end{figure}

Figure \ref{fig_adepos_cmp} summarizes a comparative study among \textit{ADIC} with all $7$ BLs that are turned on for full precision computing throughout the lifetime of a machine at different supply voltages and a case when the ADEPOS algorithm is run on \textit{ADIC} at lower supply voltage. It has been observed that when the ADEPOS algorithm runs on \textit{ADIC}, it keeps $80\%$ of the chip in inactive mode (clock gated) for $\approx 99\%$ of the lifetime based on testing with the NASA bearing dataset. This yields an energy efficiency of $0.48 pJ/OP$, an $18.5X$ and $6.96X$ reduction over the full-precision compute by \textit{ADIC} at $1.2V$ and $0.75V$ with all $7$ BLs turned on respectively. Of course, the savings obtained are entirely data dependent. As an example, we provide an example of energy savings in another hypothetical case when $80\%$ of \textit{ADIC} is in an inactive mode for around $90\%$ of the lifetime with just one BL tuned ON. In this case, energy efficiency due to ADEPOS is obtained as $0.64 pJ/OP$.

In a generic case, we can calculate the effective energy efficiency for \textit{ADIC} over the entire lifetime of a machine data as: 
\begin{equation}
E_{total} = \sum_i \alpha_{i} E(N_i) \hspace{0.1in} s.t. \hspace{0.1in} N_i \in \{1, 3, \cdots K \}
\label{eqn5a}
\end{equation}
Here, $E(N_i)$ denotes the energy dissipated by \textit{ADIC} when $N_i$ BLs are active at any point of time and $\alpha_{i}$ denotes the fraction of the total lifetime during which $N_i$ BLs are active. When \textit{ADIC} is running with full accuracy (i.e. without ADEPOS), $\alpha_{i} = 1$ for $N_i = K$ and the rest of the $\alpha_{i}$s are zero. However, for energy savings according to the ADEPOS algorithm, $\alpha_{i}$ has the largest value when $N_i = 1$ and has significantly decreasing value with the increasing number of active BLs ($N_i = 3, 5, \cdots K$) in case when \textit{ADIC} starts to flag a fault. 

When ADEPOS runs on \textit{ADIC} during the healthy state of a machine, it performs with similar accuracy as \textit{ADIC} without ADEPOS running on it and thus effectively saves the number of computes. In order to compute the energy efficiency for ADEPOS, we divide the total energy dissipated $E_{total}$ by the same number of computes required by \textit{ADIC} operating with $K$ active BLs for the same accuracy computation throughout its lifetime. As mentioned earlier, the current implementation of \textit{ADIC} has a maximum number of BLs $K$ as $7$. Hence, the ADEPOS algorithm can potentially achieve a maximum $7X$ improvement in energy saving in case of no-fault detection when only one BL is active.

\subsection{Energy Saving in Training: OPIUM-Lite}
\begin{figure}[!ht]
\centering
\includegraphics[scale=0.6]{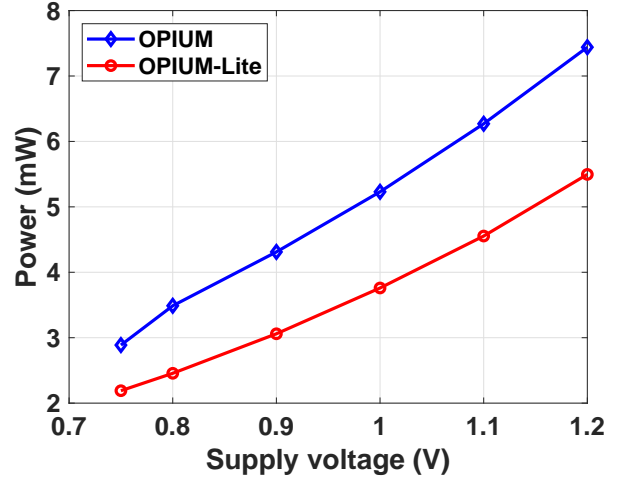}
\caption{ \textit{ADIC} training power deploying OPIUM and OPIUM-Lite algorithms at $10MHz$ ($K = 7$).}
\label{fig_trng_lite}
\end{figure}

The results presented in Fig. \ref{fig_trng_lite}, shows the power consumption during the online training phase of \textit{ADIC} leveraging OPIUM and OPIUM-Lite online learning algorithm respectively. We obtain the results for $K = 7$, at $10MHz$ clock frequency while the supply voltage $V_{dd}$ is varied from $0.75V$ to $1.2V$. We achieve an average of $42.8\%$ training energy savings when OPIUM-Lite mode is used over OPIUM.

\begin{figure}[!ht]
\centering
\includegraphics[scale=0.34]{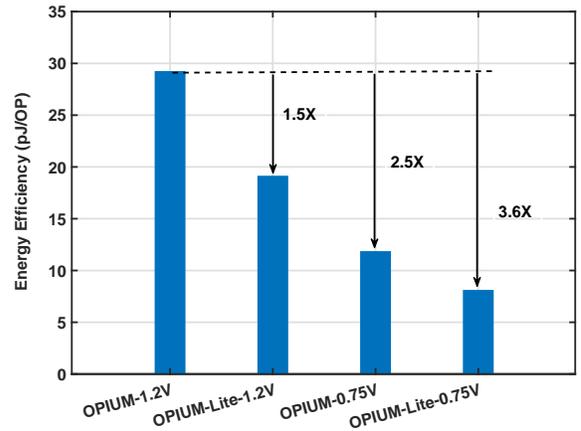}
\caption{Comparing Energy Efficiency ($pJ/OP$) for Training: OPIUM vs OPIUM-Lite.}
\label{fig_opium_lite_trng_cmp}
\vspace{-1.5em}
\end{figure}

As discussed in the previous section, this power saving is a cumulative effect due to the fact that OPIUM-Lite mode turns off the update mechanism for $\theta$ parameters in the respective memory of each BL, thus saving the dynamic power consumed by the respective memories. The energy efficiency computed for the OPIUM mode is $11.87 pJ/OP$, while OPIUM-Lite exhibits a better energy efficiency of $8.12 pJ/OP$ at $0.75V$ and $10MHz$. The plot in Fig. \ref{fig_opium_lite_trng_cmp} summarizes that OPIUM-Lite running at $0.75V$ mode can achieve a maximum $3.6X$ energy efficiency with respect to OPIUM operating at $1.2V$ supply. 

\subsection{Performance Evaluation in PdM}
In order to validate the online learning of \textit{ADIC} in machine health monitoring applications, we opt for the widely used NASA bearing dataset provided by the Center for Intelligent Maintenance Systems (IMS), University of Cincinnati \cite{NasaDataset}. The dataset consists of $12$ time-series run to failure bearing data and $4$ of them have been failed at the end of their life. The vibration data of the bearings are sampled at $20$KHz for $1$s duration at an interval of $10$ minutes. In order to reduce the amount of data seen by \textit{ADIC}, we extract statistical time-domain features such as (a) RMS, (b) Kurtosis, (c) Peak-Peak, (d) Crest factor, and (e) Skewness for each 20K data samples which are shown to be useful for bearing health monitoring \cite{MARTIN199567}. Although frequency-domain features are also used for machine health monitoring in the industries to a great extent \cite{873206} but are expensive in terms of the number of computes. Due to space limitation, we confined ourselves to showcase the time-series response of \textit{ADIC} on one dataset that exhibits a failure trend towards the end of its lifetime. Similar responses were obtained for all the other failing dataset, while the responses for all non-failing dataset do not show any failure.

As per the description of the NASA bearing dataset, there is no specific information regarding the time instance at which failures start to occur, except the success/failure label provided for each dataset. Hence, we determine a \textit{threshold} ($Thr$) that draws a boundary between the response indicating good health of a machine under consideration and that of a fault. Here, we assume that the machine remains healthy during its early life and degrades only after a significant lifespan. Hence, initial data from the machine are used to train the BLs.

\subsubsection{Threshold Selection}
In order to estimate a threshold value ($Thr$), we leverage a leave-one-out strategy (leave $1$ out of $12$ bearing dataset for testing, while the rest are used for training). In this method, the reconstruction errors (the difference between the expected and actual output of \textit{ADIC}), $e$, for all $11$ training bearings data are calculated as follows:
\begin{equation}
e=\norm{\mathbf{x}-\tilde{\mathbf{x}}}
\label{recons_error}
\end{equation}
where $\textbf{x}$ and $\tilde{\textbf{x}}$ denote the input and reconstructed output vectors respectively. Following this, the threshold $Thr$ is calculated according to Eq. (\ref{eqn4}):
\begin{equation}
Thr=\mu_{e}+0.1\times k\times \sigma_{e}
\label{eqn4}
\end{equation}
where $\mu_{e}$ and $\sigma_{e}$ denote the mean and standard deviation of the reconstructed errors respectively.
The calculated $Thr$ is then used to test the remaining (left out) bearing data, in order to observe the response being a true fault or not. We iterate this approach for each of $12$ bearing data to complete the cross-validation. 

\begin{figure}[!ht]
\centering
\begin{subfigure}[b]{0.23\textwidth}
\centering
\includegraphics[scale=0.31]{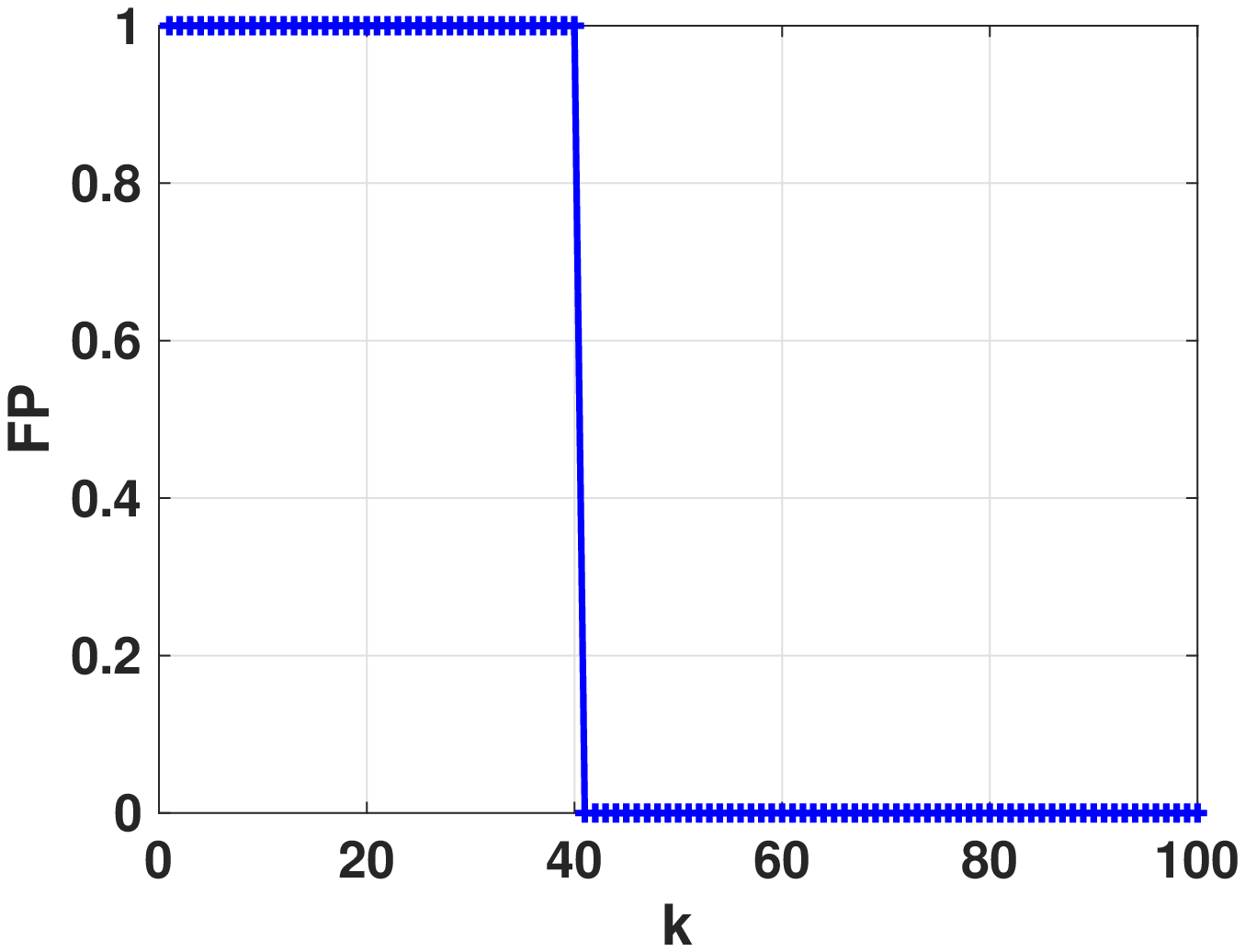}
\caption{}
\end{subfigure}
\begin{subfigure}[b]{0.23\textwidth}
\centering
\includegraphics[scale=0.31]{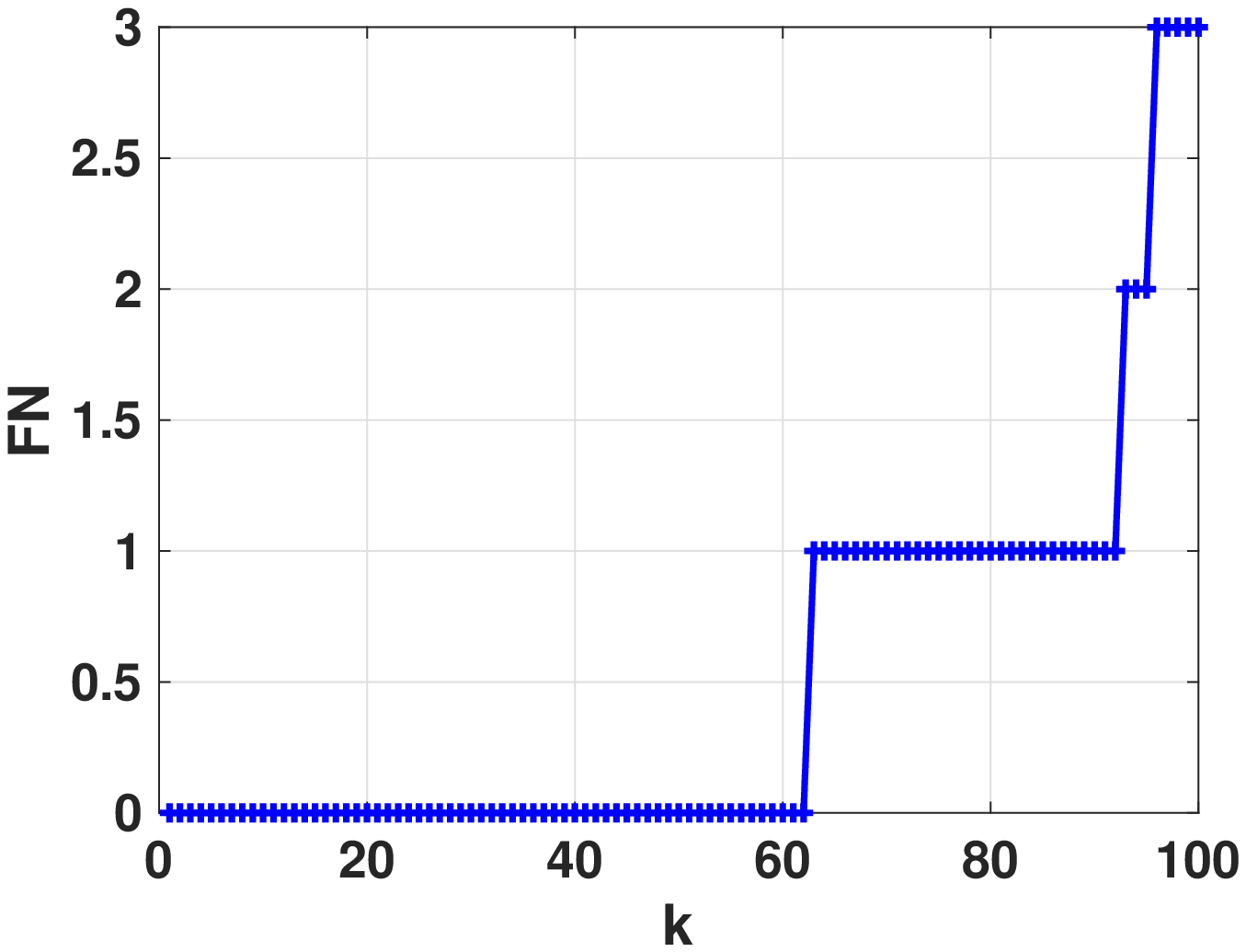}
\caption{}
\end{subfigure}
\caption{Cross-validation for selecting the optimum value of $k$: (a) FP vs $k$, and (b) FN vs $k$}
\label{fig_cval}
\end{figure}

For a good estimate of $k$ used in Eq. (\ref{eqn4}), the values of $k$ are swept from $10$ to $100$ and both false positives ($FP$) and false negatives ($FN$) are observed. In this work, we define true positive (TP) as good health and true negative (TN) for fault. The results of this exercise are presented in Fig. \ref{fig_cval} and it shows the optimum value of $k$ lies between $40$ to $60$. We choose the optimum value as $50$ which guarantees that no faulty state of a machine is classified as a healthy state ($FP = 0$) and a good machine is flagged as a faulty machine ($FN = 0$). Although the latter is not a critical situation, it may lead to over-maintenance of the machine.

\begin{figure*}[!h]
\centering
\begin{subfigure}[b]{0.47\textwidth}
\centering
\includegraphics[scale=0.5]{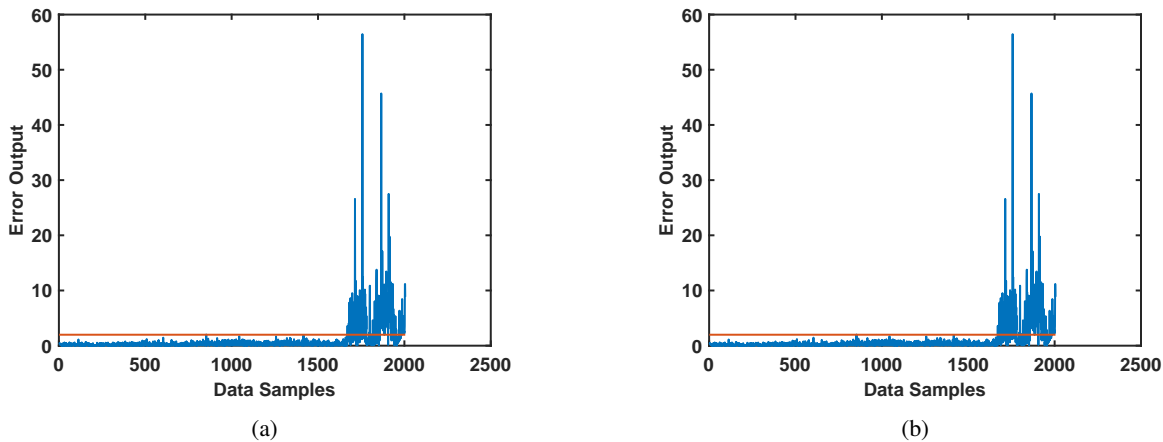}
\caption{}
\end{subfigure}
\begin{subfigure}[b]{0.47\textwidth}
\centering
\includegraphics[scale=0.5]{images/Bad_data_0p75.eps}
\caption{}
\end{subfigure}
\caption{Reconstruction error of \textit{ADIC} for a faulty machine from NASA bearing dataset: (a) OPIUM at $0.75V$, (b) OPIUM-Lite at $0.75V$ demonstrating both algorithms are successful in detecting fault at approximately the same time.}
\label{fig_err_resp}
\end{figure*}
\subsubsection{Time Series Response}
Figure \ref{fig_err_resp} presents a set of reconstructed error responses of \textit{ADIC} for the time-series data pertaining to a faulty machine given in the NASA bearing dataset. The responses for both OPIUM and OPIUM-Lite modes are obtained at $0.75V$ and $10MHz$. The x-axis in the figure represents the sample count of the data, while the y-axis represents the reconstruction errors from \textit{ADIC}. The threshold ($Thr$) is plotted in red color in each of these plots and represents the classification boundary for anomaly detection. It can be seen that the response for the vibration data pertaining to the faulty machine is crossing the threshold line approximately around the sample count of $1700$, while the response for a good machine never crosses the threshold.

It is to be noted that both the modes used the healthy data prevailing during the early stage of the machine's life, similar to the sub-threshold error response for the first few hundred data samples in these plots. We also note that these responses are quite similar to our Matlab simulation results presented in Fig. \ref{fig_err_matlab} for the same dataset.  The test error crosses the threshold near the end of the life of bearing data. The plots are slightly different due to a) Fig. \ref{fig_err_matlab} is plotted in semi-log scale b) data is scaled and  converted to 7-bit signed integer format for the processing in ADIC.


\subsection{Comparison with Existing Works}
In Table \ref{tab:ml_comp}, we present a comparative study of this work, with and without ADEPOS, and other recent published neural network integrated circuits \cite{Vsze17,Moons17,Gonug18,Quest18,Ardakani18,Wmough17}. The proposed \textit{ADIC} implements an ensemble of two layer fully connected (FC) network that is most similar to \cite{Wmough17,Gonug18} while most other works have implemented convolutional networks (CNN). Moreover, we exploit the random weights in our first layer to replace weight memory with a PRBS generator. Additionally, most of these works depend on offline learning while the proposed \textit{ADIC} implements online learning. The energy efficiency in our approach comes from usage of approximate computing at several layers while using a standard digital flow that does not rely on special memory macros for in-memory computing. It can be seen that energy efficiency of our proposed approach is competitive with respect to these recent published results. Furthermore, we compare ADIC with the microprocessor-based anomaly detector ~\cite{adepos_j}. The performance of ADIC is an order of magnitude better in terms of energy efficiency.   
\begin{table*}[!ht]
  \scriptsize
  \centering
  \caption{Comparing \textit{ADIC} (with ADEPOS) with some existing ML Co-Processors}
    \begin{tabular}{|c|c|c|c|c|c|c|c|c|c|c|}
    \hline
      & \multicolumn{2}{|c|}{This work (ADIC)} & \multicolumn{8}{|c|}{Existing Works} \\ \cline{2-4}
      \hline
     {Parameters} & w/o ADEPOS & w/ ADEPOS &  {\cite{Vsze17}}  & \multicolumn{2}{|c|}{\cite{Moons17}} &  {\cite{Gonug18}}    &   {\cite{Quest18}}    & {\cite{Ardakani18}} & {\cite{Wmough17}} & {\cite{adepos_j}}\\  \cline{2-11}
     \hline
    {Technology} & \multicolumn{3}{|c|}{65nm} & \multicolumn{2}{|c|}{40nm} & 65nm  & 40nm  & 65nm & 28nm & 65nm \\  \cline{2-10}
     \hline
    {Network Type} & \multicolumn{2}{|c|}{SLFN} & {Conv-DNN} & \multicolumn{2}{|c|}{Conv-DNN} & {SVM} & {Conv-DNN} & {Conv-DNN} &  FC-DNN & SLFN\\  \cline{2-10}
     \hline
    {Chip Measurements} & \multicolumn{7}{|c|}{Yes} & {No} & {Yes}  & Yes\\  \cline{2-10}
     \hline
    {Core Area ($mm^2$)} & \multicolumn{2}{|c|}{2.56} & 12.25 & \multicolumn{2}{|c|}{2.4} & 1.44  & 121.55 & 1.77 & 5.76 & 0.662\\  \cline{2-10}
     \hline
    {Supply Voltage ($V$)} & 0.75  & 0.75  & 0.82  & 0.6   & 1.1   & 1.0    & 1.1   & 1.0 & 0.77 & 0.75\\  \cline{2-10}
     \hline
    {Training Method} & \multicolumn{2}{|c|} {Online} & Offline & \multicolumn{2}{|c|}{Offline} & Online & Offline & Offline & Offline & Offline\\  \cline{2-10}
     \hline
    {On-Chip Memory ($KB$)} & \multicolumn{2}{|c|}{22} & 181.5 & \multicolumn{2}{|c|}{148} & 16  & 7680  & 43 & 1152 & -\\  \cline{2-10}
     \hline
    {Frequency ($MHz$)} & \multicolumn{2}{|c|}{10} & 200   & 12    & 204   & 1000  & 300   & 400 & 667 & 15.5\\  \cline{2-10}
     \hline
    {Energy Efficiency ($TOPS/W$)} & \textbf{0.298} & \textbf{2.076} & 0.123 & 0.900  & 0.270   & 3.120  & 7.490 (1.960)  & 0.277 & 1.852 & 0.03 \\  \cline{2-10}
     \hline
    \end{tabular}%
  \label{tab:ml_comp}%
\end{table*}%


\section{Conclusion}
\label{sec:conclusion}
In this paper, we present an anomaly detection integrated circuit (\textit{ADIC}), a machine learning co-processor based on extreme learning machines (ELM) algorithmic framework. \textit{ADIC} consists of an ensemble of one-class classification (OCC) engines for anomaly detection in machine health monitoring, using majority voting. This design implements an online pseudoinverse update method (OPIUM), learning from in-situ data coming from the sensor nodes attached to the machines under health monitoring and is fabricated using $65$nm CMOS process. \textit{ADIC} enables $\approx 3.6$ times energy savings during training by using a combination of an approximate training algorithm (OPIUM Lite) and DVFS. During inference, it can save $\approx 18.8$ times energy over a baseline model exploiting approximate computing algorithm ADEPOS combined with DVFS. The experiments are performed with the NASA bearing dataset, popularly used for evaluating predictive maintenance algorithms; however, the presented ADIC can be used for other anomaly detection algorithms as well. 

\section*{Acknowledgment}
This work was conducted within the Delta-NTU Corporate Lab for Cyber-Physical Systems with funding support from Delta Electronics Inc. and the National Research Foundation (NRF) Singapore under  Corp-Lab@University Scheme.

\balance
\bibliographystyle{IEEEtran}
\bibliography{BibFiles/IC2_bibliography}

\balance
\vspace{-1em}
\begin{IEEEbiographynophoto}{Bapi Kar}
(S'16-M’17) received his Bachelor of Instrumentation Engineering from Jadavpur University India in 2002, and PhD from Indian Institute of Technology Kharagpur, India in 2017. He was the recipient of University Gold Medal from Jadavpur University in 2002. He worked as a design and verification engineer in VLSI industry for more than 6 years. 

Currently, he is a postdoctoral research fellow at Nanyang Technological University Singapore since October 2017. His current research interests include Low power integrated circuit (IC) design for machine learning applications, VLSI Physical Design Automation Algorithms and Design for Manufacturability (DFM) issues in nanometer technologies.
\end{IEEEbiographynophoto}
\vspace{-3em}

\begin{IEEEbiographynophoto}{Pradeep Kumar Gopalakrishnan}
(M’01-SM'10) received B.Tech in Electrical Engineering from the University of Kerala in 1989 and M.Tech in Electronics Design and Technology from the Indian Institute of Science in 1997. He has over 25 years of experience in the industry, mainly in ASIC and embedded systems design. He worked in companies such as Philips, Broadcom, Institute of Microelectronics (A*STAR), Siemens and Xilinx before joining NTU. 

He is currently pursuing a PhD degree at Nanyang Technological University, Singapore. His research interests include low-power Machine Learning architectures and Neuromorphic Systems.
\end{IEEEbiographynophoto}
\vspace{-3em}

\begin{IEEEbiographynophoto}{Sumon Kumar Bose}
(S’18) received Bachelor of Engineering in Electronics and Telecommunication Engineering from Jadavpur University, Kolkata in 2012. He has five years of VLSI industry experience working in companies like Cypress Semiconductor, xSi Semiconductor and Texas Instruments. He worked on several Analog circuit design related projects. 

Currently, he is a PhD student at Nanyang Technological University Singapore since January 2017. His research interests include Low power analog IC, and Machine Learning based Hardware Design.
\end{IEEEbiographynophoto}

\vspace{-3em}
\begin{IEEEbiographynophoto}{Mohendra Roy}
(M’17) received his Ph.D. in Electronic and Information Engineering from Korea University, South Korea in 2016. He did his Masters in BioElectronics as well as Physics from Tezpur University, India in 2008 and 2006 respectively. He received Gold Medal from Tezpur University in 2008. Dr. Roy received the Korea University Graduate Achievement Award in 2016, IEEE Student Paper Contest award (Seoul Section) in 2014 and Outstanding Paper award in Biochip 2014 Fall Conference in South Korea. He served as a session chair at IEEE SSCI 2018 conference (in Feature Analysis track). 

Currently, he is a assistant professor at School of Technology and Science, PDPU, India. Prior to that he was a post-doctoral research fellow at Delta-NTU corporate lab, Nanyang Technological University. His research interests include AI, Bio-Photonics, and Bio-Sensors.
\end{IEEEbiographynophoto}


%
\vspace{-3em}
\begin{IEEEbiographynophoto}{Arindam Basu}
(M’10-SM'17) received the B.Tech. and the M.Tech. degrees in electronics and electrical communication engineering from IIT Kharagpur in 2005, and the MS degree in mathematics and the Ph.D. degree in electrical engineering from the Georgia Institute of Technology, Atlanta, in 2009 and 2010, respectively. He joined Nanyang Technological University, Singapore in 2010, where he currently holds a tenured Associate Professor position. He received the Prime Minister of India Gold Medal in 2005 from IIT Kharagpur. 

His research interests include bio-inspired neuromorphic circuits, non-linear dynamics in neural systems, low-power analog IC design, and programmable circuits and devices. He was a Distinguished Lecturer of the IEEE Circuits and Systems Society for the 2016–2017 term. He received the Best Student Paper Award from the Ultrasonics symposium in 2006, the best live demonstration at ISCAS 2010 and a finalist position in the best student paper contest at ISCAS 2008. He also received the MIT Technology Reviews inaugural TR35@Singapore Award in 2012 for being among the top 12 innovators under the age of 35 in Southeast Asia, Australia, and New Zealand. He was a Guest Editor for two special issues in the IEEE Transactions on Biomedical Circuits and Systems for selected papers from ISCAS 2015 and BioCAS 2015. He is serving as a Corresponding Guest Editor for the special issue on low-power, adaptive neuromorphic systems: devices, circuit, architectures and algorithms in the IEEE Journal on Emerging Topics in Circuits and Systems. He is currently an Associate Editor of the IEEE Sensors Journal, the IEEE Transactions on Biomedical Circuits and Systems, and the Frontiers in Neuroscience. 
\end{IEEEbiographynophoto}

\end{document}